\documentclass[runningheads]{llncs}

\usepackage{graphicx}
\usepackage[misc]{ifsym}
\usepackage{easyReview}

\usepackage{graphicx}
\usepackage{amsfonts}
\usepackage{graphicx}
\usepackage{epstopdf}
\usepackage{algorithmic}
\usepackage[toc,page]{appendix}

\usepackage{enumitem}
\usepackage{listings}
\usepackage{color} 
\definecolor{mygreen}{RGB}{28,172,0} 
\definecolor{mylilas}{RGB}{170,55,241}

\newcommand{\R}{\ensuremath{\mathbb{R}}}

\newcommand{\Ll}{{L}}

\newcommand{\Dd}{{D}}
\newcommand{\Ww}{{W}}

\newcommand{\eps}{\varepsilon}

\def\abs#1{\left|#1\right|} %
\def\sym{\mathrm{sym}}

\def\vol{\mathop{\rm vol}\nolimits}
\usepackage{tikz,pgfplots}
\usepackage{textcomp}

\usepackage{amsmath}
\usepackage{enumitem}
\usepackage{graphicx}
\usepackage{array,multirow,graphicx}

\usepackage{comment}
\usepackage{subcaption}
\def\pedro#1{\textcolor{black}{#1}}

\def\pedroCameraReady#1{\textcolor{black}{#1}}

   \newcommand{\old}[1]{}  %

\newcommand{\oldd}[1]{}  %

\usepackage[algo2e,ruled,linesnumbered]{algorithm2e}

\usepackage[numbers,sort&compress]{natbib}

\def\Smatrix{S}%

\usepackage{adjustbox}

\usepackage{tabularx}

\begin{document}
\title{ Node Classification for Signed Social Networks Using Diffuse Interface Methods }
\titlerunning{Node Classification for Signed Networks Using Diffuse Interface Methods}
%
\author{\Letter{Pedro Mercado}\inst{1} \and
Jessica Bosch\inst{2} \and
Martin Stoll\inst{3}}
\authorrunning{P. Mercado et al.}
%
\tocauthor{Pedro~Mercado (University of T\"ubingen),
Jessica~Bosch (The University of British Columbia),
Martin~Stoll (Technische Universit\"at Chemnitz),
}%
\toctitle{Node Classification for Signed Social Networks Using Diffuse Interface Methods}
\institute{
Department of Computer Science, University of T\"ubingen, Germany \and
Department of Computer Science, The University of British Columbia, Canada \and
Faculty of Mathematics, Technische Universit\"at Chemnitz, Germany
}
\maketitle              
%
\begin{abstract}
 Signed networks contain both positive and negative kinds of interactions like friendship and enmity. The task of node classification in non-signed graphs has proven to be beneficial in many real world applications, yet extensions to signed networks remain largely unexplored. In this paper we introduce the first analysis of node classification in signed social networks via diffuse interface methods based on the Ginzburg-Landau functional together with different extensions of the graph Laplacian to signed networks. We show that blending the information from both positive and negative interactions leads to performance improvement in real signed social networks, consistently outperforming the current state of the art.
\end{abstract}

\section{Introduction}

Signed graphs are graphs with both positive and negative edges, where positive edges encode relationships like friendship and trust, and negative edges encode conflictive and enmity interactions.
Recently, signed graphs have received an increasing amount of attention due to its capability to encode
interactions that are not covered by unsigned graphs \pedroCameraReady{or multilayer graphs~\cite{Mercado:2018:powerMean,Serafino:2018,Tang:2009:CMG:1674659.1677085,zhou2007spectral,Tudisco:2018:NL}}, which mainly encode interactions based on similarity and trust.

While the analysis of unsigned graphs follows a long-standing and well established tradition~\cite{Luxburg:2007:tutorial,Newman:2006a,BerF12},
the analysis of signed graphs can be traced back to~\cite{Cartwright:1956:Structural,Harary:1954:Notion}, in the context of social balance theory, further generalized in~\cite{Davis:1967:Clustering} by introducing the concept of a $k$-balance signed graph: a signed graph is $k$-balanced if the set of nodes can be partitioned into $k$ disjoint sets such that inside the sets there are only positive relationships, and between \pedroCameraReady{different sets} only negative relationships. 
\pedroCameraReady{A related concept is constrained clustering~\cite{basu2008constrained}, where must-links and cannot-links are constraints indicating if certain pairs of nodes should be assigned to the same or different clusters.}

Recent developments of signed graphs have been guided by the concept of $k$-balance, leading to a diverse paradigm of applications, including:
clustering~\cite{Doreian:2009:Partitioning,Chiang:2012:Scalable,Kunegis:2010:spectral,Mercado:2016:Geometric,sedoc:2017a,kirkley2018balance,Cucuringu:2019:SPONGE,Cucuringu:MBO,Mercado:2019:signedPowerMean}, 
edge prediction~\cite{kumar2016wsn,Leskovec:2010:PPN,LeFalher:2017a},
node embeddings~\cite{derr:2018:signedGraphConvolutionalNetwork,Kim:2018:SRL:3178876.3186117,Wang:2017:signedNetworkEmbedding,Yuan:2017:SNE},
node ranking~\cite{Chung:2013,Shahriari:2014}, 
node classification~\cite{tang2016node},
and many more. See~\cite{tang2015survey,gallier2016spectral} for a recent survey on the topic.
\color{black}
One task that remains largely unexplored is the task of node classification in signed networks.

The problem of
node classification in graphs is a semi-supervised learning problem
where the goal is to improve classification performance by taking into account both labeled and unlabeled observations~\cite{Zhu:Book2009,Chapelle:2010:SL}, being a particular case graph-based semi-supervised learning.
%

The task of graph-based classification methods on unsigned graphs is a fundamental problem with many application areas~\cite{Zhou:2003:LLG:2981345.2981386,Zhu:2003:SLU:3041838.3041953,Belkin:2004}.
A technique that has recently been proposed with very promising results utilizes techniques known from partial differential equations in materials science and combines these with graph based quantities (cf.~\cite{BerF12}). In particular, the authors 
in~\cite{BerF12}
use diffuse interface methods that are derived from the Ginzburg--Landau energy~\cite{AllC79,BE93,GNS1999,TC94}. 
These methods have been used in image inpainting where a damaged region of an image has to be restored \old{to incorporate} given information about the undamaged image parts. 
%
In the context of node classification in graphs, the undamaged part of an image corresponds to labeled nodes, whereas the damaged part corresponds to unlabeled nodes to be classified based on the information of the underlying graph structure of the image and available labeled nodes.
With this analogy, one can readily use results from  \cite{BerEG07} for the classification problem on graphs. While the materials science problems are typically posed in an infinite-dimensional setup, the corresponding problem in the 
graph-based
classification problem uses the graph Laplacian. This technique has shown great potential and has recently been extended to different setups \old{\cite{GarMBFP14,MerGBFP14,BosKS16}}\pedro{\cite{BosKS16,GarMBFP14,MerGBFP14}}.

\pedroCameraReady{
Our contributions are as follows:
we study the problem of node classification in signed graphs by developing a natural extension of diffuse interface schemes of Bertozzi and Flenner\pedro{~\cite{BerF12}}, based on different signed graph Laplacians. 
To the best of our knowledge this is the first study of node classification in signed networks using diffuse interface schemes.
A main challenge when considering the application of diffuse interface methods to signed networks is the availability of several competing \pedroCameraReady{signed} graph Laplacians and how the method's performance depends on the chosen \pedroCameraReady{signed} graph Laplacian, hence we present a thorough comparison of our extension based on existing signed graph Laplacians.
%
%
Further, we show the effectivity of our approach against state of the art approaches by performing extensive experiments on real world signed social networks.
}

The paper is structured as follows. We first introduce the tools needed from graphs and how they are extended to signed networks. We study the properties of several different signed Laplacians. We then introduce a diffuse interface technique in their classical setup and illustrate how signed Laplacians can be used within the diffuse interface approach. 
This is then followed by 
\pedroCameraReady{ 
numerical experiments in real world signed networks.
}
\\
\\
\textbf{Reproducibility}: Our code is available at \texttt{https://github.com/melopeo/GL}
\section{Graph information and signed networks}

We now introduce the Laplacian for unsigned graphs followed by particular versions used for signed graphs.
\subsection{Laplacians for unsigned graphs}
%
In this section we introduce several graph Laplacians, which are the main tools for our work.
Let $G=(V,W)$ be an undirected graph with node set $V=\{v_1,\ldots,v_n\}$ of size $n=\abs{V}$ and adjacency matrix $W\in\R^{n\times n}$ with non-negative weights, i.e., $w_{ij}\geq 0$. 

In the case where a graph presents an assortative configuration, i.e.  edge weights of the adjacency matrix $W$ represent similarities (the larger the value of $w_{ij}$ the larger the similarity of nodes the $v_i$ and $v_j$), then the Laplacian matrix is a suitable option for graph analysis, as the eigenvectors corresponding to the $k$-smallest eigenvalues convey an embedding into $\R^k$ such that similar nodes are close to each other~\cite{Luxburg:2007:tutorial}. The Laplacian matrix and its normalized version are defined as:
\begin{equation*}
 L = D - W, \quad\qquad\,\, L_\sym = D^{-1/2}LD^{-1/2}
\end{equation*}
%
%
%
%
%
where $D\in\R^{n\times n}$ is a diagonal matrix with $D_{ii} = \sum_{i=1}^n w_{ij}$. 
Observe that $L_\sym$ can be further simplified to $L_\sym=I-\Dd^{-1/2}\Ww\Dd^{-1/2}$.
Both Laplacians $L$ and $L_\sym$ are symmetric positive semi-definite, and the multiplicity of the eigenvalue zero is equal to the number of connected components in the graph $G$.
\\

For the case where a graph presents a dissasortative configuration, i.e. edges represent dissimilarity (the larger the value of $w_{ij}$ the more dissimilar are the nodes $v_i$ and $v_j$), then the signless Laplacian is a suitable option, as the eigenvectors corresponding to the $k$-smallest eigenvalues provide an embedding into $\R^k$ such that dissimilar nodes are close to each other~\cite{Desai:1994:characterization,Liu2015,Mercado:2016:Geometric}.
The signless Laplacian matrix and its normalized version are defined as:
\begin{equation*}
 Q = D + W, \quad\qquad\,\, Q_\sym = D^{-1/2}QD^{-1/2}
\end{equation*}
Observe that $Q_\sym$ can be further simplified to $Q_\sym=I+\Dd^{-1/2}\Ww\Dd^{-1/2}$.
Both Laplacians $Q$ and $Q_\sym$ are symmetric positive semi-definite, with smallest eigenvalue equal to zero if and only if there is a bipartite component in $G$.
\\
\\
We are now ready to introduce the corresponding Laplacians for the case where both positive and negative edges are present, to later study its application to node classification in signed graphs.
%

\subsection{Laplacians for signed graphs}\label{subsection:LaplaciansForSignedGraphs}
\pedroCameraReady{We are now ready to present different signed graph Laplacians.} We give a special emphasis on the particular notion of a cluster that each signed Laplacian aims to identify. This is of utmost importance, since this will influence the classification performance of our proposed method.

Signed graphs are useful for the representation of positive and negative interactions between a fixed set of entities. We define a signed graph to be a pair $G^\pm=(G^+,G^-)$ where $G^+=(V,W^+)$ and $G^-=(V,W^-)$ contain positive and negative interactions respectively,
between the same set of nodes $V$, with symmetric adjacency matrices $W^+$ and $W^-$. 
For the case where a single adjacency matrix $W$ contains both positive and negative edges, one can obtain the signed adjacency matrices by the relation $W^+_{ij} = \max (0,W_{ij})$ and $W^-_{ij} = -\min (0,W_{ij})$.

\textbf{Notation}:
we denote the positive, negative and absolute degree diagonal matrices as
$D^+_{ii}=\sum_{j=1}^n W^+_{ij}$, $D^-_{ii}=\sum_{j=1}^n W^-_{ij}$ and $\bar{D}=D^++D^-$;
the Laplacian and normalized Laplacian of positive edges as
$L^+~=~D^+ - W^+$, and $L^+_\sym~=~(D^+)^{-1/2} L^+ (D^+)^{-1/2}$;
and for negative edges
$L^-~=~D^- - W^-$, and $L^-_\sym = (D^-)^{-1/2} L^- (D^-)^{-1/2}$,
together with the signless Laplacian for negative edges
$Q^-~=~D^- + W^-$, and $Q^-_\sym~=~(D^-)^{-1/2} Q^- (D^-)^{-1/2}$.

A fundamental task in the context of signed graphs is to find a partition of the set of nodes~$V$ such that inside \pedroCameraReady{the} clusters there are \pedroCameraReady{mainly} positive edges, and between \pedroCameraReady{different} clusters there are \pedroCameraReady{mainly} negative edges. This intuition corresponds to the concept of $k$-balance of a signed graph, which can be traced back to~\cite{Davis:1967:Clustering}:
\textit{
A signed graph is $\mathbf{k}$\textbf{-balanced} if the set of vertices can be partitioned into $k$ sets such that within the subsets there are only positive edges, and between them only negative. 
}

Based on the concept of $k$-balance of a signed graph, 
several extensions of the graph Laplacian to signed graphs have been proposed, each of them aiming to bring a $k$-dimensional embedding of the set of nodes $V$ through the eigenvectors corresponding to the $k$-smallest eigenvalues, such that positive edges
keep nodes close to each other, and negative edges
push nodes apart.

Examples of extensions of the graph Laplacian to signed graphs are the signed ratio Laplacian and its normalized version~\cite{Kunegis:2010:spectral}, defined as
\begin{equation*}
 \Ll_{SR}=\bar{\Dd}-\Ww, \qquad \Ll_{SN}={I}-\bar{\Dd}^{-1/2}\Ww\bar{\Dd}^{-1/2}
\end{equation*}
Both Laplacians are positive semidefinite. Moreover, they have a direct relationship to the concept of 2-balance of a graph, as their smallest eigenvalue is equal to zero if and only if the corresponding signed graph is 2-balanced. Hence, the magnitude of the smallest eigenvalue tells us how far a signed graph is to be 2-balanced.
In~\cite{Kunegis:2010:spectral} it is further observed that the quadratic form $x^T \Ll_{SR} x$ is related to the discrete signed ratio cut optimization problem:
\begin{equation*}
 \min_{C\subset V} \left(\, 2\text{cut}^+(C,\overline{C}) + \text{assoc}^-(C) + \text{assoc}^-(\overline{C}) \, \right) \left( \frac{1}{\abs{C}} + \frac{1}{\abs{\overline{C}}} \right)
\end{equation*}
where 
$\overline{C}=V\backslash C$, 
$\text{cut}^+(C,\overline{C}) = \sum_{i\in C, j\in \overline{C}} W^+_{ij}$ counts the number of positive edges between clusters, and 
$\text{assoc}^-(C) = \sum_{i\in C, j\in C} W^-_{ij}$
counts the number of negative edges inside cluster $C$ (similarly for $\text{assoc}^-(\overline{C})$). Therefore we can see that the first term counts the number of edges that keeps the graph away from being 2-balanced, while the second term enforces a partition where both sets are of the same size.
\newline
\pedroCameraReady{Inspired by the signed ratio cut,}
the balance ratio Laplacian and its normalized version are defined as follows~\cite{Chiang:2012:Scalable}:
\begin{equation*}
 L_{BR} = D^+ - W^+ + W^-,  \qquad L_{BN} = \bar{D}^{-1/2} L_{BR} \bar{D}^{-1/2}, 
\end{equation*}
Observe that these Laplacians need not be positive semi-definite, i.e. they potentially have negative eigenvalues. Further, the eigenvectors corresponding to the smallest eigenvalues of $L_{BR}$ are inspired by the following discrete optimization problem:
\begin{equation*}
 \min_{C\subset V} \left(
 \frac{\text{cut}^+(C,\overline{C}) + \text{assoc}^-(C)}{\abs{C}}
 +
 \frac{\text{cut}^+(C,\overline{C}) + \text{assoc}^-(\overline{C})}{\abs{\overline{C}}}
 \right)
\end{equation*}
A further proposed approach, based on the optimization of some sort of ratio of positive over negative edges (and hence denoted SPONGE) is expressed through the following generalized eigenvalue problem and its normalized version~\cite{Cucuringu:2019:SPONGE}:
\begin{equation*}
 (L^+ + D^-)v = \lambda (L^- + D^+)v\, ,
 \qquad
  (L^+_\sym + I)v = \lambda (L^-_\sym + I)v
\end{equation*}
which in turn are inspired by the following discrete optimization problem
\begin{equation*}
 \min_{C\subset V} \left(
 \frac{ \text{cut}^+(C,\overline{C}) + \vol^-(C) }{\text{cut}^-(C,\overline{C}) + \vol^+(C)}
 \right)
\end{equation*}
where $\vol^+(C)=\sum_{i\in C} d^+_i$ and $\vol^-(C)=\sum_{i\in C} d^-_i$.
Observe that the normalized version corresponds to the eigenpairs of $L_{\text{SP}}:=(L^-_\sym + I)^{-1}(L^+_\sym + I)$.
Finally, based on the observation that the signed ratio Laplacian can be expressed as the sum of the Laplacian and signless Laplacian of positive and negative edges, i.e. $L_{SR} = L^+ + Q^-$, in~\cite{Mercado:2016:Geometric} the arithmetic and geometric mean of Laplacians are introduced:
\begin{equation*}
  L_{AM} = L_\sym^+ + Q_\sym^-,  \qquad L_{GM} = L_\sym^+ \# Q_\sym^-\,.
\end{equation*}
%

Observe that different clusters are obtained from different signed Laplacians.
This becomes clear as different clusters are obtained as solutions from the related discrete optimization problems above described. In the following sections we will see that different signed Laplacians induce different classification performances in the context of graph-based semi-supervised learning on signed graphs.
\color{black}

\section{Diffuse interface methods}\label{section:DiffuseInterfaceMethods}

Diffuse interface methods haven proven to be useful in the field of materials science~\cite{AllC79,BE93,CahH58,EllS10,GNSW08} with applications to phase separation, biomembrane simulation~\cite{WanD08}, image inpainting~\cite{BerEG07,BosKSW14} and beyond. In~\cite{BerF12} it is shown that diffuse interface methods provide a novel perspective to the task of graph-based semi-supervised learning. 
These methods are commonly based on the minimization of the Ginzburg-Landau (\textbf{GL}) functional, which itself relies on a suitable graph Laplacian.
Let $S\in\R^{n\times n}$ be a positive semi-definite matrix. We define the \textbf{GL} functional for graph-based semi-supervised learning as follows:
\begin{equation}
\label{gl1Pedro}
E_S(u):=\frac{\eps}{2} u^{T}S u+\frac{1}{4\eps}\sum_{i=1}^n(u_i^2-1)^2+\sum_{i=1}^n\frac{\omega_i}{2}(f_i-u_i)^{2}\,,
\end{equation}
where $f_i$ contains the class labels of previously annotated nodes.
%

Observe that this definition of the GL functional for graphs depends on a given positive semi-definite matrix $S$. For the case of non-signed graphs a natural choice is the graph Laplacian (e.g. $S=L_\sym$), which yields the setting presented in~\cite{BerF12,GarMBFP14,MerGBFP14}. In the setting of signed graphs considered in this paper one can \pedroCameraReady{utilize} only the information encoded by positive edges (e.g. $S=L_\sym^+$), only negative edges (e.g. $S=Q_\sym^-$), or 
both for which a positive semi-definite signed Laplacian that blends the information encoded by both positive and negative edges is a suitable choice (e.g. $S=L_\text{SR},L_\text{SN}, L_{\text{SP}}, \text{ or } L_\text{AM}$).
%
\\
\newline
Moreover, each element of the GL functional plays a particular role:
\begin{enumerate}[topsep=-3pt,leftmargin=*]
 \item $\frac{\eps}{2} u^{T}S u$ induces smoothness and brings clustering information of the signed graph. Different choices of $S$ convey information about different clustering assumptions, as observed in Section~\ref{subsection:LaplaciansForSignedGraphs},
 \item $\frac{1}{4\eps}\sum_{i=1}^n(u_i^2-1)^2$ has minimizers with entries in $+1$ and $-1$, hence for the case of two classes it induces a minimizer $u$ whose entries indicate the class assignment of unlabeled nodes,
 \item $\sum_{i=1}^n\frac{\omega_i}{2}(f_i-u_i)^{2}$ is a fitting term to labeled nodes given \textit{a priori}, where $\omega_i=0$ for unlabeled nodes and $\omega_i=w_0$ for labeled nodes, with $w_0$ large enough (see Sec.~\ref{V20-sec:datasets} for an analysis on $w_0$.)
 \item The interface parameter $\eps>0$ allows to control the trade-off between the first and second terms: large values of $\eps$ make the clustering information provided by the matrix $\Smatrix$ more relevant, whereas small values of $\eps$ give more weight to vectors whose entries correspond to class assignments of unlabeled nodes (see Sec.~\ref{V20-sec:datasets} for an analysis on $\eps$.)
 \\
\end{enumerate}
%
Before briefly discussing the minimization of the GL functional $E_S(u)$, note that the matrix $\Smatrix$ needs to be positive semi-definite, as otherwise the $E_S(u)$ becomes unbounded below. This discards signed Laplacians like the balance ratio/normalized Laplacian introduced in section~\ref{subsection:LaplaciansForSignedGraphs}.
%
The minimization of the GL functional $E_S(u)$ in the $L^2$ 
function space
sense can be done through a gradient descent leading to a modified Allen-Cahn equation.
We employ a convexity splitting scheme (see
\cite{BerEG07,BosKSW14,BosKS16,Eyr98,luo2016convergence,SchB11,van2014mean}),
where the trick is to split $E_S(u)$ into a difference of convex functions:
\begin{itemize}[topsep=-3pt,leftmargin=*]
 \item[] \centering $E_\Smatrix(u)=E_1(u)-E_2(u)$     
\end{itemize}
with 
\begin{align*}
 E_1(u)& =\frac{\eps}{2} u^{T}\Smatrix u+\frac{c}{2}u^{T} u\,, \\
 E_2(u)&=\frac{c}{2}u^{T} u-\frac{1}{4\eps}\sum_{i=1}^n(u_i^2-1)^2-\sum_{i=1}^n\frac{\omega_i}{2}(f_i-u_i)^2
\end{align*}
where 
$E_1$ and $E_2$
are convex if $c\geq\omega_{0}+\frac{1}{\varepsilon}$; 
(see e.g. \cite{BosKS16}).
Proceeding with an implicit Euler scheme for $E_1$ and explicit treatment for $E_2$, leads to the following scheme:
\begin{equation*}
 \frac{u^{(t+1)}-u^{(t)}}{\tau} = -\nabla E_1(u^{(t+1)}) + \nabla E_2(u^{(t)})
\end{equation*}
where 
$\left( \nabla E_1(u) \right)_i = \frac{\partial E_1}{\partial{u_i}} (u) $
and
$\left( \nabla E_2(u) \right)_i = \frac{\partial E_2}{\partial{u_i}} (u) $
with $i=1,\ldots,n$,
and
$u^{(t+1)}$ (resp. $u^{(t)}$) is the evaluation of $u$ at the current (resp. previous) time-point.
This further leads to the following
\begin{align*}
\frac{u^{(t+1)}-u^{(t)}}{\tau}+\eps \Smatrix  u^{(t+1)}+cu^{(t+1)}
=cu^{(t)}
-\frac{1}{\eps}\nabla\psi(u^{(t)})
+\nabla\varphi(u^{(t)}).
\end{align*}
where
$\psi(u) = \sum_{i=1}^n(u_i^2-1)^2$ and $\varphi(u)=\sum_{i=1}^n\frac{\omega_i}{2}(f_i-u_i)^2$.
\\
Let $(\lambda_l,\phi_l)$, $l=1,\ldots,n$, be the eigenpairs of $\Smatrix$. 
By projecting terms of the previous equation onto the space generated by eigenvectors $\phi_1,\ldots,\phi_n$, we obtain
\begin{align}
&\frac{a_l-\bar{a}_l}{\tau}+\eps \lambda_l a_l+ca_l=-\frac{1}{\eps}\bar{b}_l+c\bar{a}_l+\bar{d}_l\quad \text{for }\,\, l=1,\ldots,n
\end{align}
where
scalars $\{(a_l, \bar{a}_l, \bar{b}_l, \bar{d}_l )\}_{l=1}^n$ 
are such that
\pedroCameraReady{
$
u^{(t+1)}=\sum_{l=1}^n a_l \phi_l$, 
$u^{(t)}=\sum_{l=1}^n \bar{a}_l \phi_l$, 
$\left(\left[\phi_1,\ldots,\phi_n\right]^T\nabla\psi\left(\sum_{l=1}^{n}\bar{a}_l\phi_l\right)\right)_l\!=\bar{b}_l$, $\left(\left[\phi_1,\ldots,\phi_n\right]^T\nabla\varphi\left(f-\sum_{l=1}^{n}\bar{a}_l\phi_l\right)\right)_l=\bar{d}_l$.}
 Equivalently, we can write this as
\begin{align}
\left(1+\eps\tau \lambda_l+c\tau\right)a_l=-\frac{\tau}{\eps}\bar{b}_l+(1+c\tau)\bar{a}_l+\tau\bar{ d}_l\quad \text{for }\,\, l=1,\ldots,n
\end{align}
where the update is calculated as $u^{(t+1)}=\sum_{l=1}^n a_l \phi_l$.
Once either convergence or the maximum of iterations is achieved, the estimated label of node $v_i$ is equal to $\text{sign}(u_i)$. The extension to more than two classes is briefly introduced in the appendix of this paper.
Finally, note that the eigenvectors corresponding to the smallest eigenvalues of a given Laplacian are the most informative, hence the projection above mentioned can be done with just a \pedroCameraReady{small} amount of eigenvectors. This will be further studied in the next section.

\section{Experiments}\label{V20-sec:datasets}
In our experiments we denote by $\textbf{GL}(S)$ our approached based on the Ginzburg-Landau functional defined in Eq.~\ref{gl1Pedro}. For the case of signed graphs we consider $\textbf{GL}(L_\text{SN}),\textbf{GL}(L_\text{SP})$, and $\textbf{GL}(L_\text{AM})$. 
To better understand the information relevance of different kind of interactions we further evaluate our method based only on positive or negative edges, i.e. $\textbf{GL}(L^+_\sym)$ and $\textbf{GL}(Q^-_\sym)$, respectively.

We compare with different kinds of approaches to the task of node classification: First, we consider transductive methods designed for unsigned graphs and apply them only to positive edges, namely: local-global propagation of labels~(\textbf{LGC})~\cite{Zhou:2003:LLG:2981345.2981386},
Tikhonov-based regularization~(\textbf{TK})~\cite{Belkin:2004}, and
Label Propagation with harmonic functions~(\textbf{HF})~\cite{Zhu:2003:SLU:3041838.3041953}.

We further consider two methods explicitly designed for the current task: \textbf{DBG}~\cite{goldberg07a}~based on a convex optimization problem adapted for negative edges, and \textbf{NCSSN}~\cite{tang2016node} a matrix factorization approach tailored for social signed networks.

\color{black}


%

\textbf{Parameter setting.}
The parameters of our method are set as follows, unless otherwise stated:
the fidelity parameter $\omega_0=10^3$, the interface parameter $\varepsilon=10^{-1}$, the convexity parameter $c=\frac{3}{\varepsilon}+\omega_0$, time step-size $dt=10^{-1}$, maximum number of iterations $2000$, stopping tolerance $10^{-6}$. 
\pedroCameraReady{Parameters of state of the art approaches are set as follows:
for LGC we set $\alpha=0.99$ following~\cite{Zhou:2003:LLG:2981345.2981386},
for TK we set $\gamma=0.001$ following~\cite{Belkin:2004},
for DBG we set $\lambda_1=\lambda_2=1$, and for
NCSSN we set $(\lambda=10^{-2},\alpha=1,\beta=0.5,\gamma=0.5)$ following~\cite{tang2016node}.
We do not perform cross validation in our experimental setting
due to the large execution time in some of the benchmark methods here considered.
Hence,
}
%
in all experiments we report the \textit{average classification accuracy} out of 10 runs, where for each run we take a different sample of labeled nodes of same size.
\\
\\
\color{black}
\begin{table}[t!]
\centering
\begin{adjustbox}{width=1\textwidth}
\addtolength{\tabcolsep}{2.5pt}    
\begin{tabular}{c|ccc|ccc|ccc|}
	  \cline{2-10}
           & \multicolumn{3}{c|}{\textbf{Wikipedia RfA}} & \multicolumn{3}{c|}{\textbf{Wikipedia Elections}} & \multicolumn{3}{c|}{\textbf{Wikipedia Editor}} \\
           \cline{2-10}
           & $G^+$     & $G^-$     & $G^\pm$   & $G^+$      & $G^-$     & $G^\pm$   & $G^+$      & $G^-$      & $G^\pm$    \\
           \cline{1-10}
\multicolumn{1}{|c|}{$\#$ nodes} & 3024      & 3124      & 3470      & 1997       & 2040      & 2325      & 17647      & 14685      & 20198      \\
\multicolumn{1}{|c|}{$+$ nodes}  & 55.2$\%$  & 42.8$\%$  & 48.1$\%$  & 61.3$\%$   & 47.1$\%$  & 52.6$\%$  & 38.5$\%$   & 33.5$\%$   & 36.8$\%$   \\
\hline
\multicolumn{1}{|c|}{$\#$ edges} & 204035    & 189343    & 215013    & 107650     & 101598    & 111466    & 620174     & 304498     & 694436     \\
\multicolumn{1}{|c|}{$+$ edges}  & 100$\%$   & 0$\%$     & 78.2$\%$  & 100$\%$    & 0$\%$     & 77.6$\%$  & 100$\%$    & 0$\%$      & 77.3$\%$  \\
\hline
\end{tabular}
\addtolength{\tabcolsep}{1.0pt}   
\end{adjustbox}
\vspace{5pt}
\caption{Dataset statistics of largest connected components of $G^+$, $G^-$ and $G^\pm$.}                                                                        
\label{table:dataStatistics}
\vspace{-20pt}
\end{table}
%
%
%
\vspace{-30pt}
\subsection{Datasets}
We consider three different real world networks: wikipedia-RfA~\cite{snapnets}, wikipedia-Elec~\cite{snapnets}, and Wikipedia-Editor~\cite{Yuan:2017:SNE}. 
%
Wikipedia-RfA and Wikipedia-Elec are datasets of editors of Wikipedia that request to become administrators, where any Wikipedia member may give a supporting, neutral or opposing vote. From these votes we build a signed network for each dataset, where a positive (resp. negative) edge indicates a supporting (resp. negative) vote by a user and the corresponding candidate. 
The label of each node in these networks is given by the output of the corresponding request: positive (resp. negative) if the editor is chosen (resp. rejected) to become an administrator. 

Wikipedia-Editor is extracted from the UMD Wikipedia dataset \cite{Kumar2015Vews}.
The dataset is composed of vandals and benign editors of Wikipedia. There is a positive (resp. negative) edge between users if their co-edits belong to the same (resp. different) categories. Each node is labeled as either benign (positive) or vandal (negative).

In the following experiments we take the largest connected component of either $G^+$, $G^-$ or $G^\pm$,
depending on the method in turn:
for LGC, TK, HF, and \textbf{GL}($L^+_\sym$) we take the largest connected component of $G^+$,
for \textbf{GL}($Q^-_\sym$) we take the largest connected component of $G^-$, and
for the remaining methods
we take the largest connected component of $G^\pm$.

In Table~\ref{table:dataStatistics} we show statistics of the corresponding \pedro{largest} connected components \old{for}\pedro{of} each dataset: all datasets present a larger proportion of positive edges than of negative edges in the corresponding signed network $G^\pm$, i.e. at least $77.3\%$ of edges are positive in all datasets. \pedro{Further, the distribution of positive and negative node labels is balanced, except for Wikipedia-Editor where the class of positive labels is between $33.5\%$ and $38.5\%$ of nodes.}


\begin{table}[t]
\centering
\begin{adjustbox}{width=1\textwidth}
\addtolength{\tabcolsep}{2.5pt}    
\begin{tabular}{|l|cccc|cccc|cccc|}
\cline{2-13}
 \multicolumn{1}{ c|}{}  & \multicolumn{4}{c|}{\begin{tabular}[l] {@{}c@{}} \rotatebox[origin=c]{0} \centering \textbf{Wikipedia}\\ \textbf{RfA}\end{tabular}} & \multicolumn{4}{c|}{\begin{tabular}[l] {@{}c@{}} \rotatebox[origin=c]{0} \centering \textbf{\color{white}\color{black}Wikipedia\color{white}\color{black}}\\ \textbf{Elections}\end{tabular}} & \multicolumn{4}{c|}{\begin{tabular}[l] {@{}c@{}} \rotatebox[origin=c]{0} \centering \textbf{Wikipedia}\\ \textbf{Editor}\end{tabular}}  \\
  \hline
 \multicolumn{1}{ |c|}{Labeled nodes}  & \color{white}\color{black} $1\%$  & \color{white}\color{black} $5\%$    & \color{white}\color{black} $10\%$    & \color{white}\color{black} $15\%$    & \color{white}\color{black} $1\%$     & \color{white}\color{black} $5\%$    & \color{white}\color{black} $10\%$    & \color{white}\color{black} $15\%$& \color{white}\color{black} $1\%$     & \color{white}\color{black} $5\%$    & \color{white}\color{black} $10\%$    & \color{white}\color{black} $15\%$  \\
   \cline{2-13}
\hline
LGC($L^+$)& 0.554 & 0.553 & 0.553 & 0.553 & 0.614 & 0.614 & 0.613 & 0.613 & 0.786 & 0.839 & 0.851 & 0.857 \\                                  
%
\pedroCameraReady{
TK($L^+$)
}& 0.676 & 0.697 & 0.681 & 0.660 & 0.734 & 0.763 & 0.742 & 0.723 & 0.732 & 0.761 & 0.779 & 0.791
\\                               
%
HF($L^+$)& 0.557 & 0.587 & 0.606 & 0.619 & 0.616 & 0.623 & 0.637 & 0.644 & 0.639 & \textbf{0.848} & \textbf{0.854} & \textbf{0.858} \\                  
%
\textbf{GL}($L^+_{\textrm{sym}}$) & 0.577 & 0.564 & 0.570 & 0.584 & 0.608 & 0.622 & 0.626 & 0.614 & 0.819 & 0.759 & 0.696 & 0.667 \\
\hline\hline                                                                                                     
DGB & 0.614 & 0.681 & 0.688 & 0.650 & 0.648 & 0.602 & 0.644 & 0.609 & 0.692 & 0.714 & 0.721 & 0.727 \\                
NCSSN & 0.763 & 0.756 & 0.745 & 0.734 & 0.697 & 0.726 & 0.735 & 0.776 & 0.491 & 0.533 & 0.559 & 0.570 \\               
\textbf{GL}($Q^-_{\textrm{sym}}$) & 0.788 & 0.800 & 0.804 & 0.804 & 0.713 & 0.765 & 0.764 & 0.766 & 0.739 & 0.760 & 0.765 & 0.770 \\
\textbf{GL}($L_{\textrm{SP}}$) & 0.753 & 0.761 & 0.763 & 0.765 & 0.789 & 0.793 & 0.797 & 0.798 & 0.748 & 0.774 & 0.779 & 0.779 \\              
\textbf{GL}($L_{\textrm{SN}}$) & 0.681 & 0.752 & 0.759 & 0.764 & 0.806 & 0.842 & 0.851 & 0.852 & \textbf{0.831} & 0.841 & 0.846 & 0.847 \\   
\textbf{GL}($L_{\textrm{AM}}$) & \textbf{0.845} & \textbf{0.847} &\textbf{ 0.848} &\textbf{ 0.849} & \textbf{0.879} & \textbf{0.885} & \textbf{0.887} & \textbf{0.887} & 0.787 & 0.807 & 0.814 & 0.817 \\   
\hline                      
\end{tabular}
\addtolength{\tabcolsep}{1pt}    
\end{adjustbox}
\vspace{5pt}
\caption{
Average classification accuracy with different amounts of labeled nodes. 
Our method \textbf{GL}($L_{\textrm{SN}}$) and \textbf{GL}($L_{\textrm{AM}}$) performs best among transductive methods for signed graphs, and outperforms all methods in two out of three datasets.
}                                                                        
\label{table:numEigenvectors_20} 
\vspace{-20pt}
\end{table}
\subsection{Comparison of Classification Performance}
In Table~\ref{table:numEigenvectors_20} we first compare our method $\textbf{GL}(S)$ with competing approaches when the amount of labeled nodes is fixed to $1\%,5\%,10\%$ and $15\%$.
%
We can see that among methods for signed graphs, our approach with $\textbf{GL}(L_\text{SN})$ and $\textbf{GL}(L_\text{AM})$ performs best. 
Moreover, in two out of three datasets our methods based on signed graphs present the best performance,
whereas for the dataset Wikipedia-Editor the unsigned graph method HF performs best. Yet, we can observe that the performance gap with our method $\textbf{GL}(L_\text{SN})$ is of at most one percent.
%
%
%
%
Overall we can see that the classification accuracy is higher when the signed graph is taken, in comparison to the case where only either positive or negative edges are considered. This suggests that merging the information encoded by both positive and negative edges leads to further improvements.

In the next section we evaluate the effect on classification performance of different amounts of labeled nodes.

\color{black}
\begin{figure}[t!]
 \centering
 \includegraphics[width=0.26\textwidth,trim=115 45 90 40,clip]{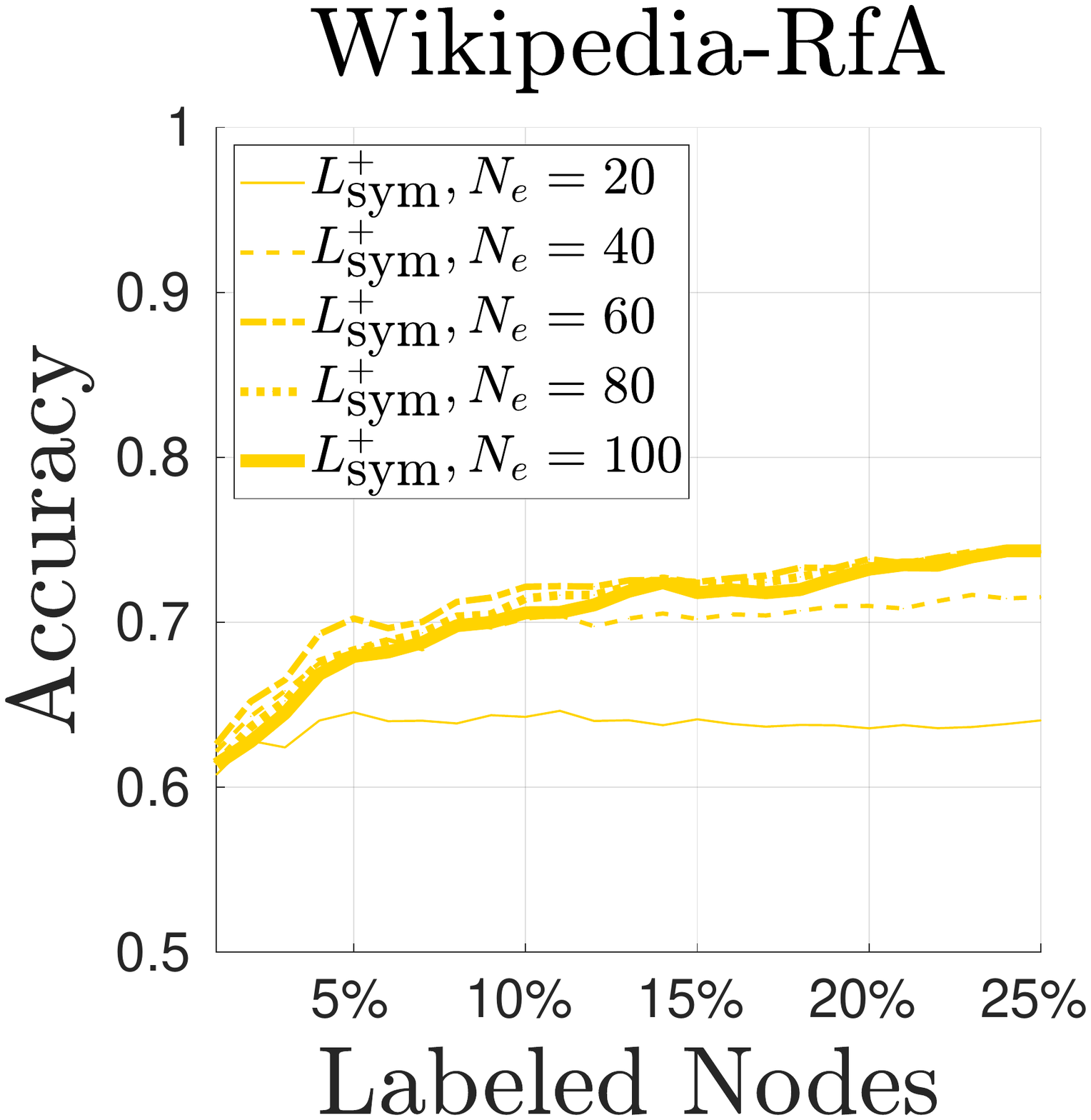}
 \includegraphics[width=0.26\textwidth,trim=115 45 90 40,clip]{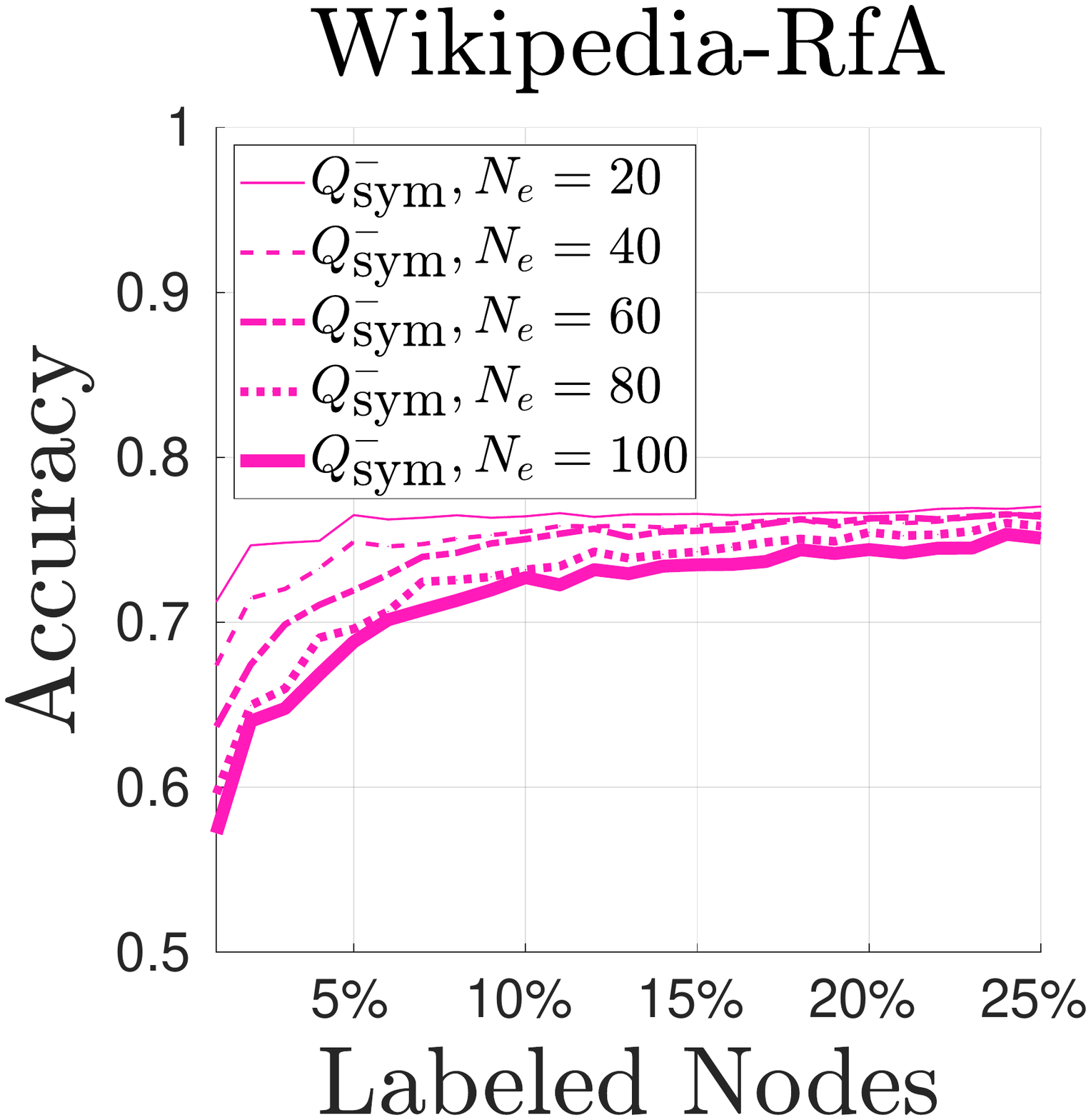}
 \includegraphics[width=0.26\textwidth,trim=115 45 90 40,clip]{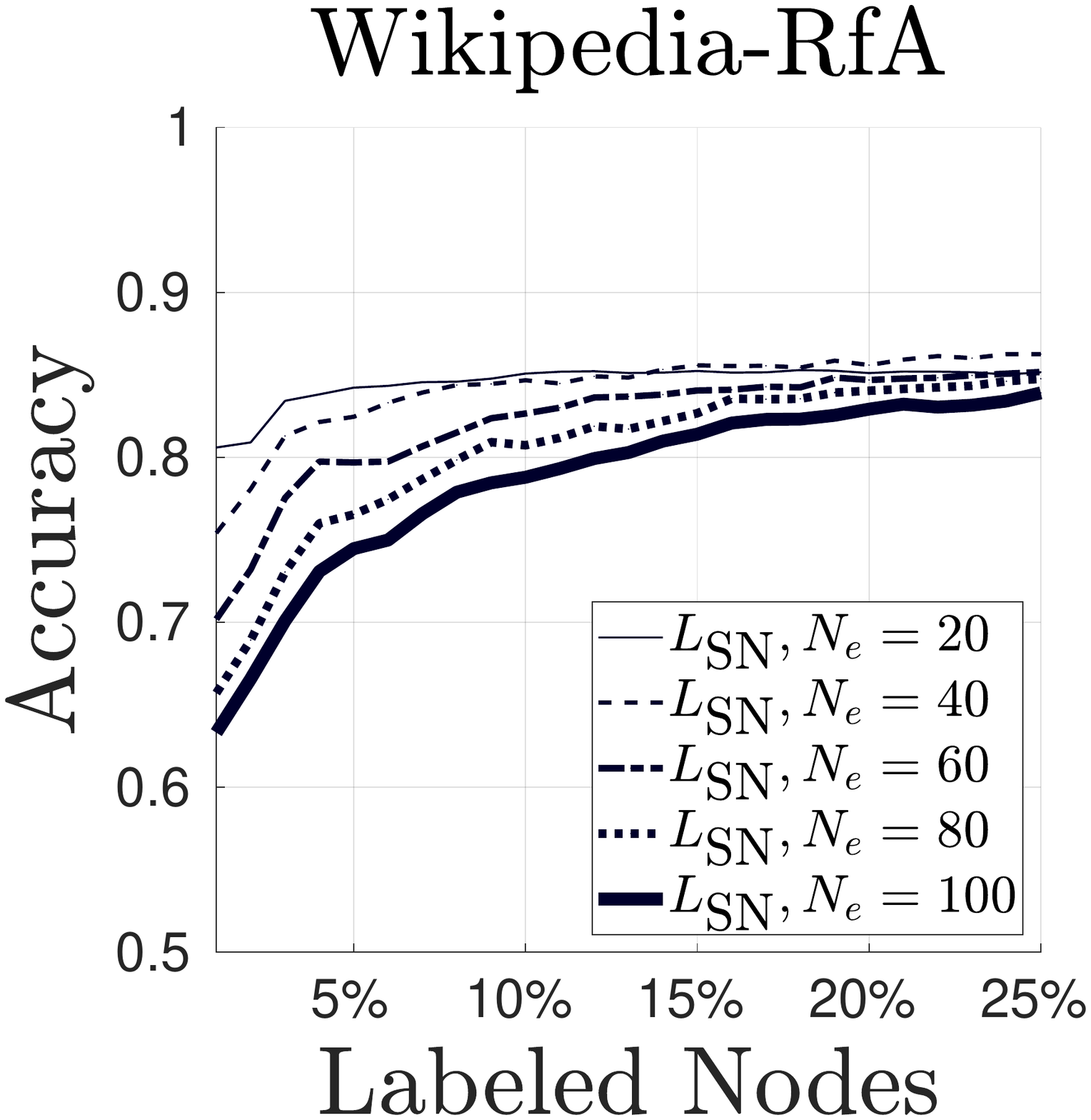}
 \includegraphics[width=0.26\textwidth,trim=115 45 90 40,clip]{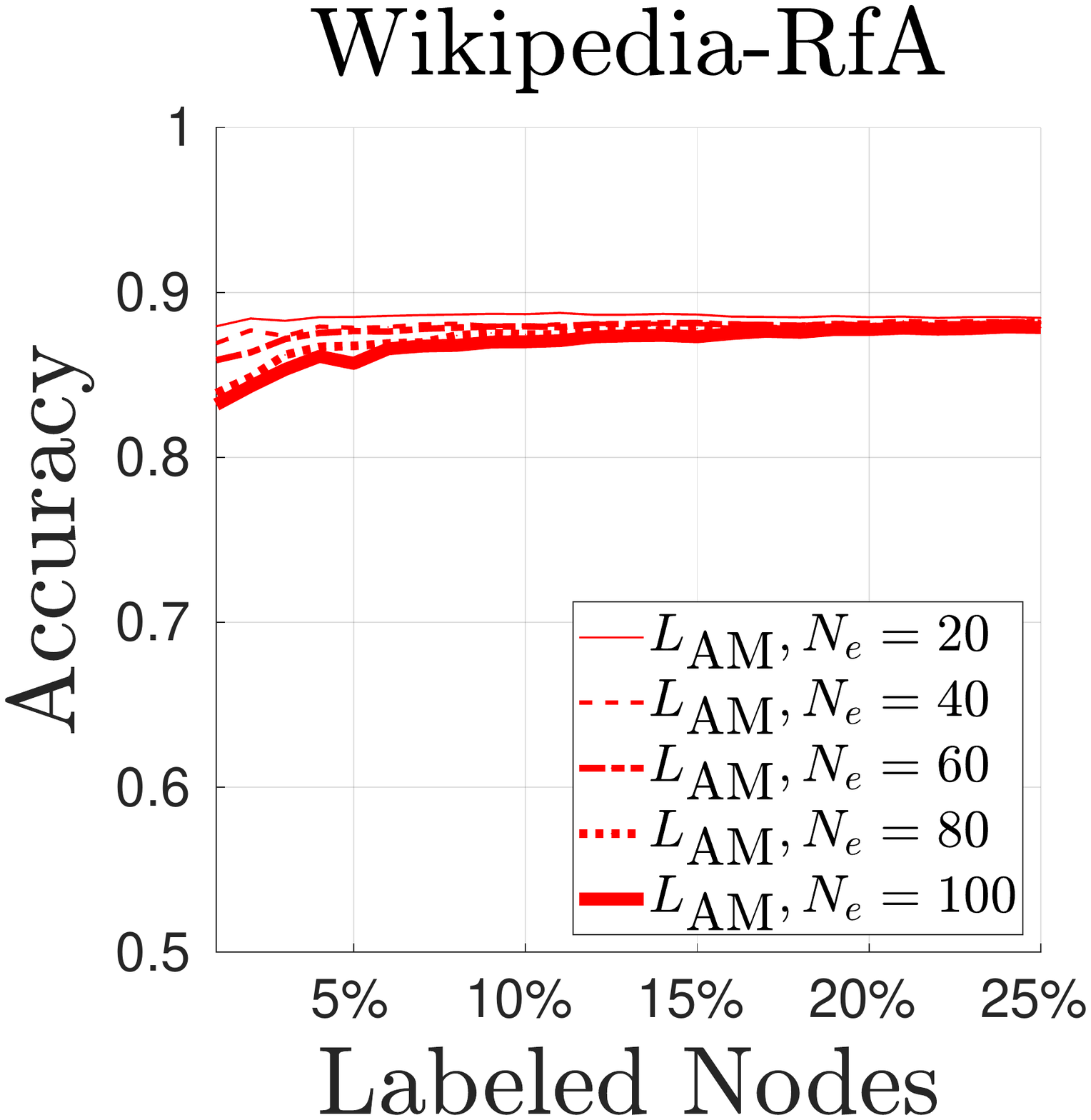}\hfill
 \\
 
  \vspace{2mm}
%
%
%
 \includegraphics[width=0.26\textwidth,trim=115 45 90 40,clip]{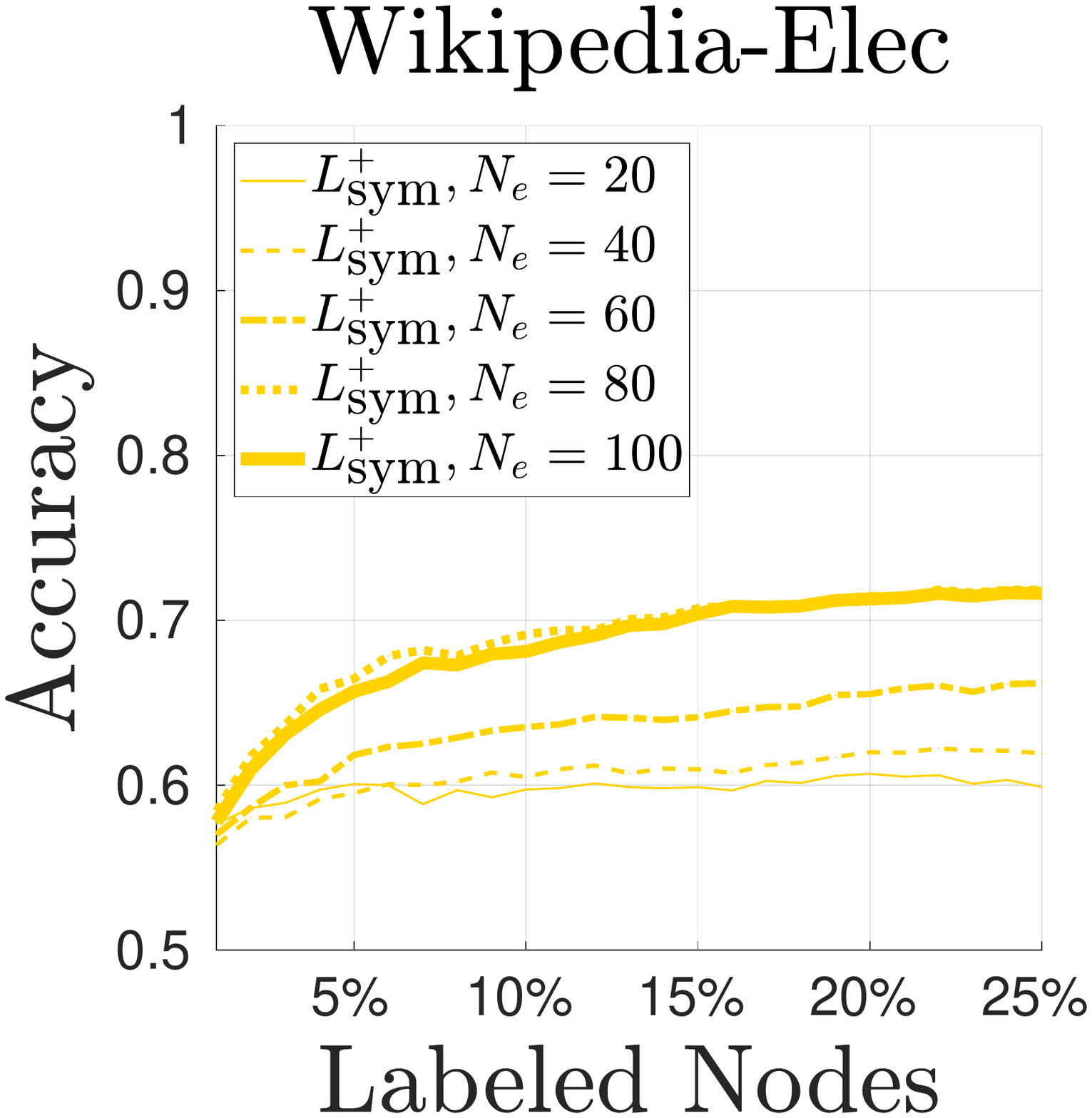}
 \includegraphics[width=0.26\textwidth,trim=115 45 90 40,clip]{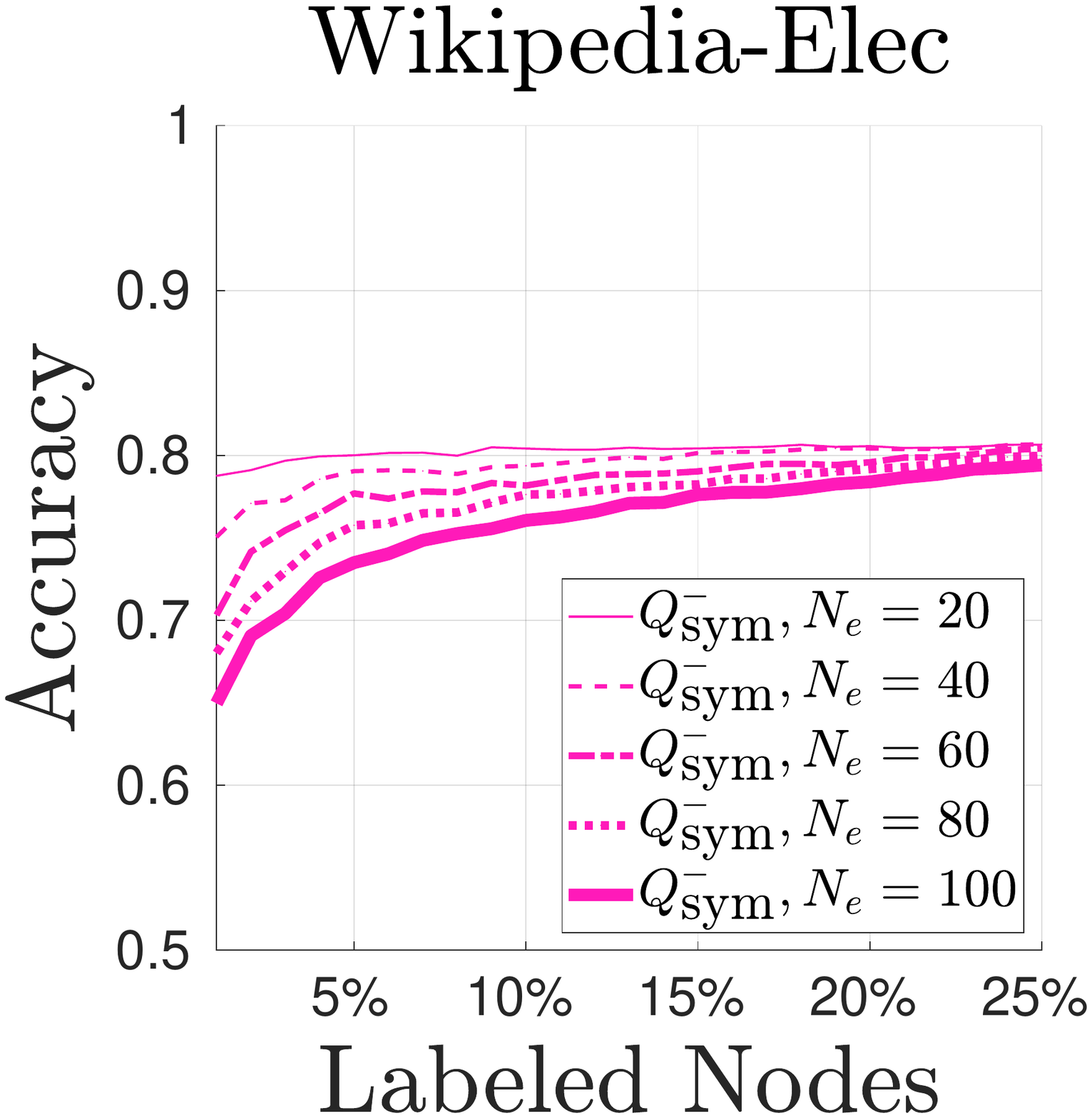}
 \includegraphics[width=0.26\textwidth,trim=115 45 90 40,clip]{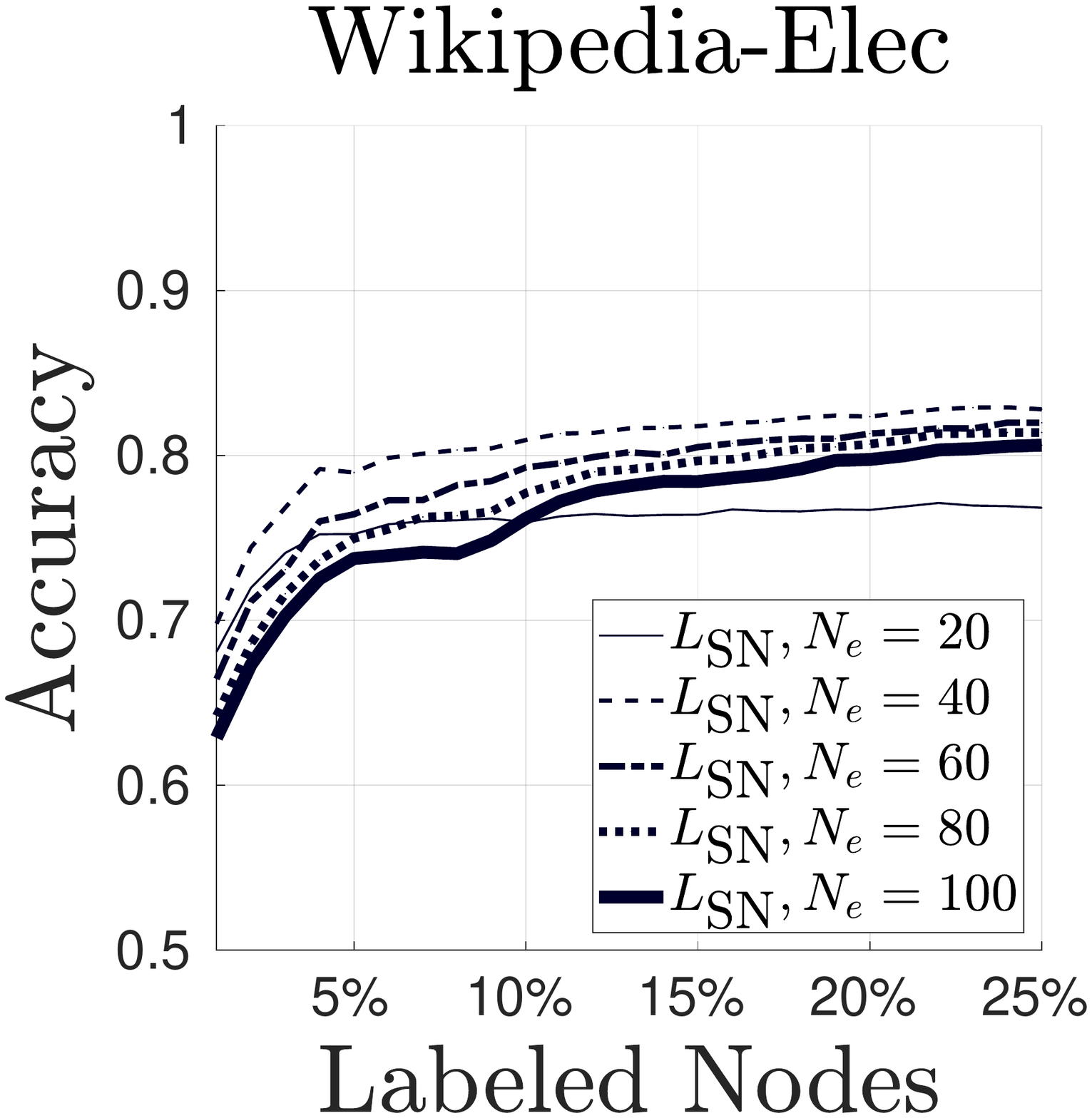}
 \includegraphics[width=0.26\textwidth,trim=115 45 90 40,clip]{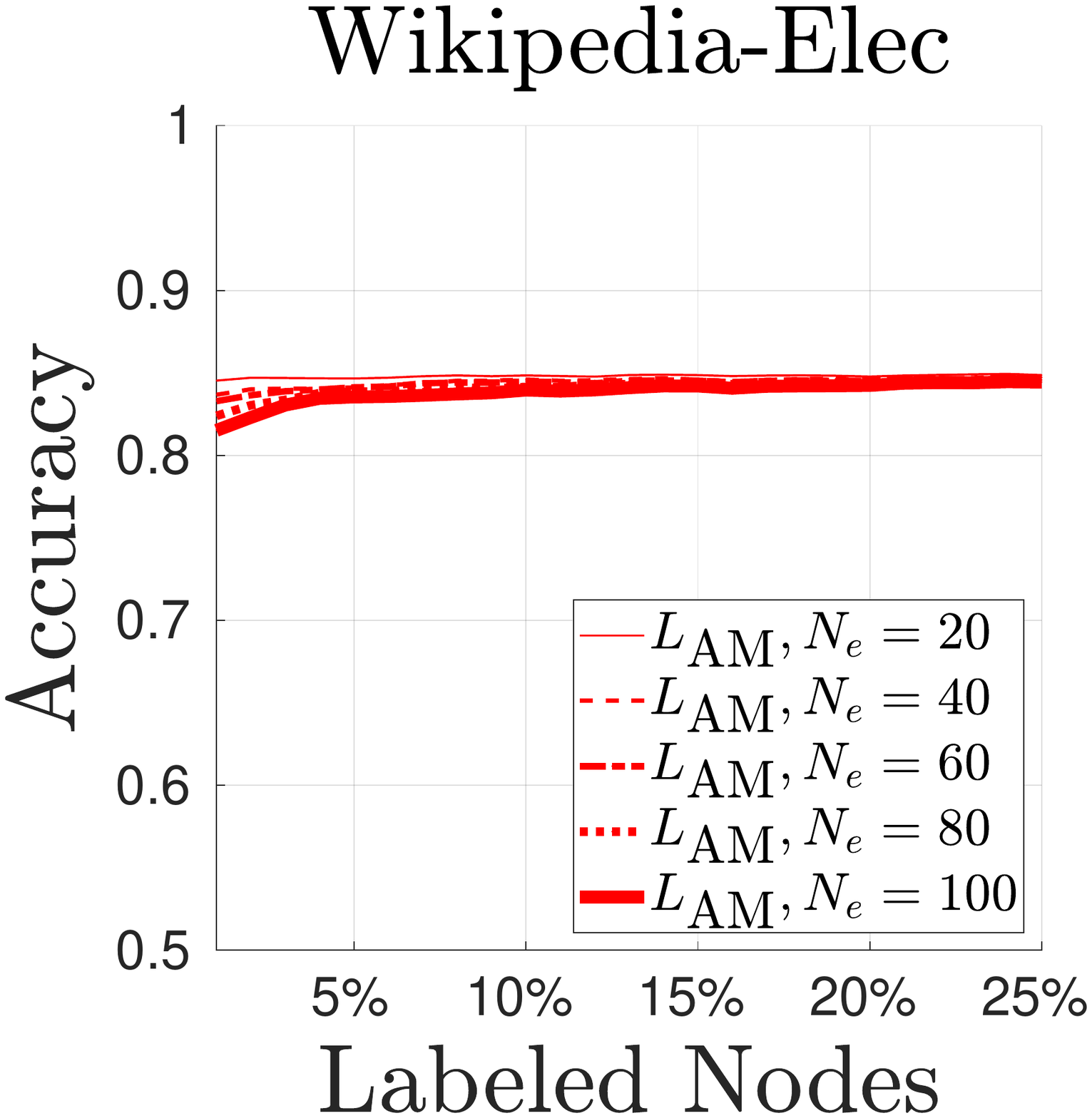}\hfill
 \\
 \vspace{2mm}
%
%
%
%
 \includegraphics[width=0.26\textwidth,trim=115 45 90 40,clip]{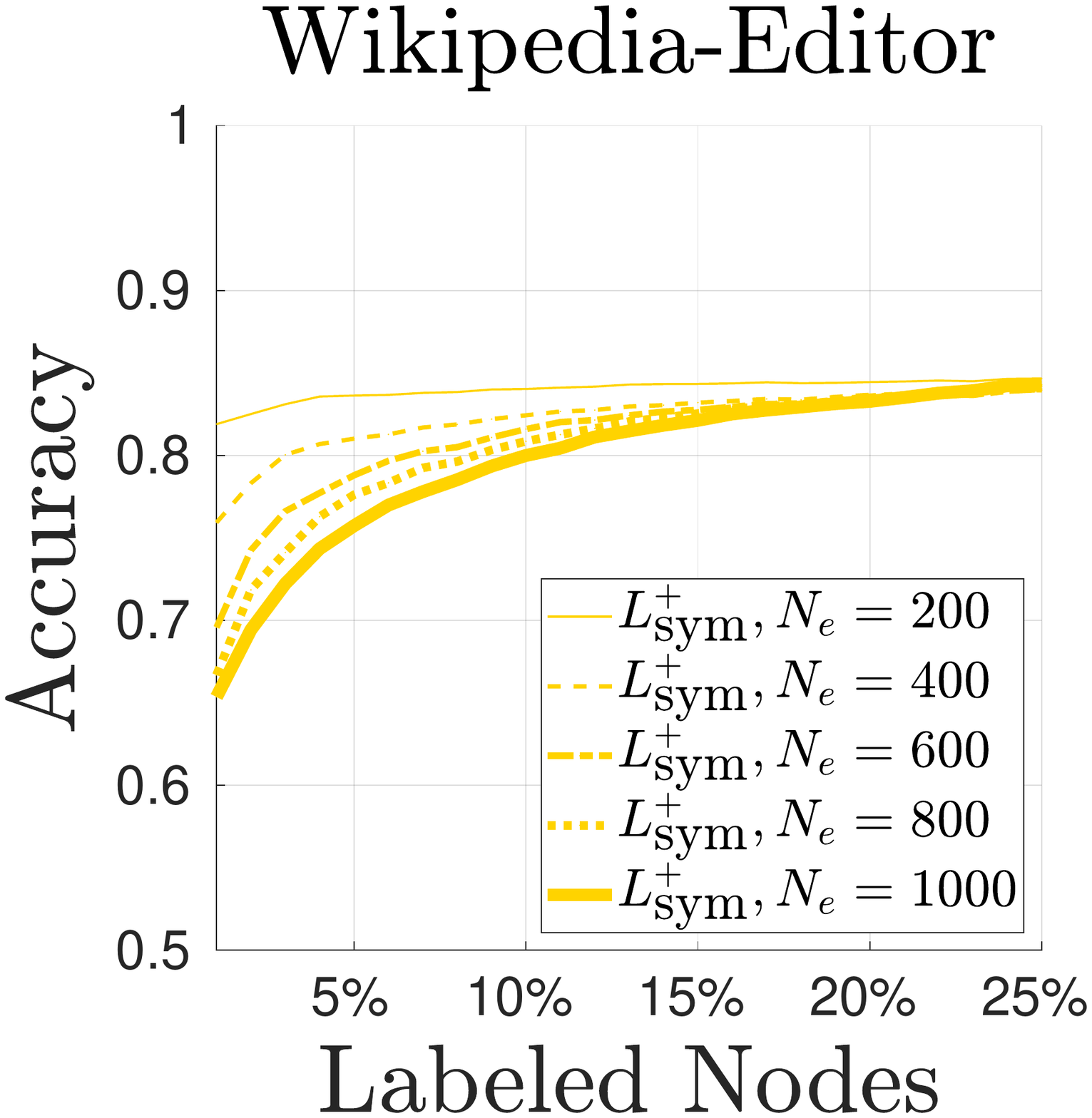}
 \includegraphics[width=0.26\textwidth,trim=115 45 90 40,clip]{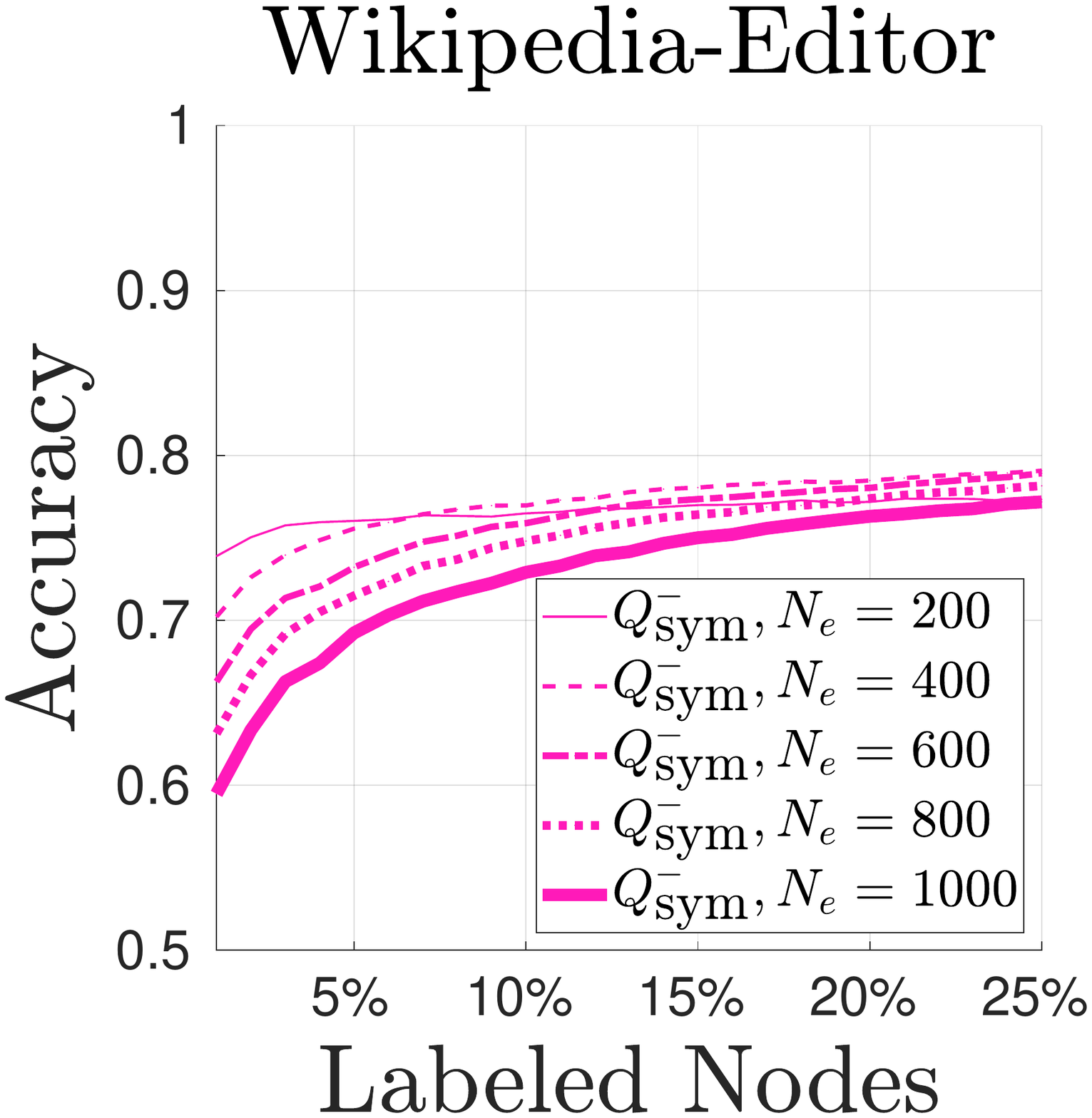}
 \includegraphics[width=0.26\textwidth,trim=115 45 90 40,clip]{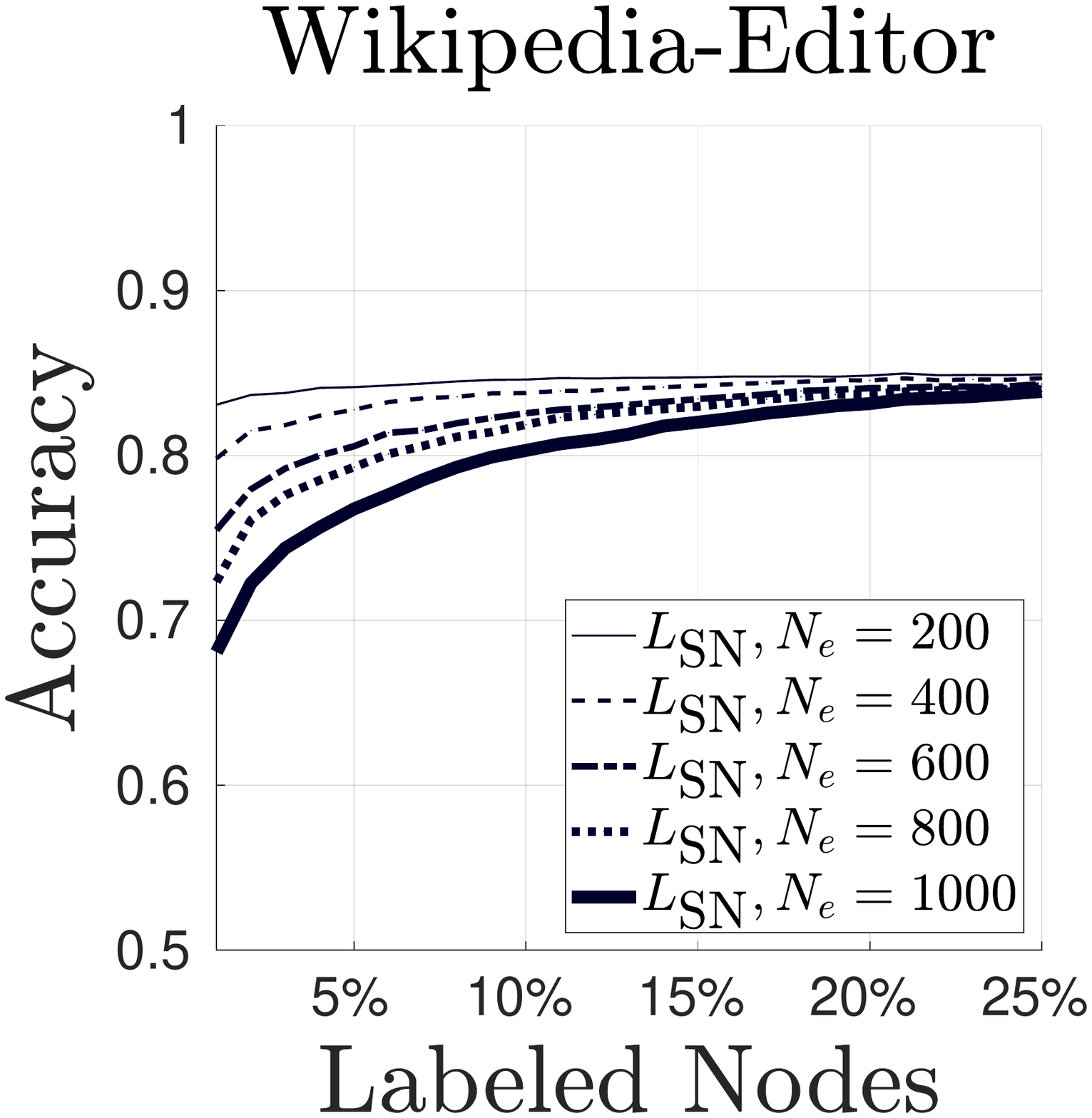}
 \includegraphics[width=0.26\textwidth,trim=115 45 90 40,clip]{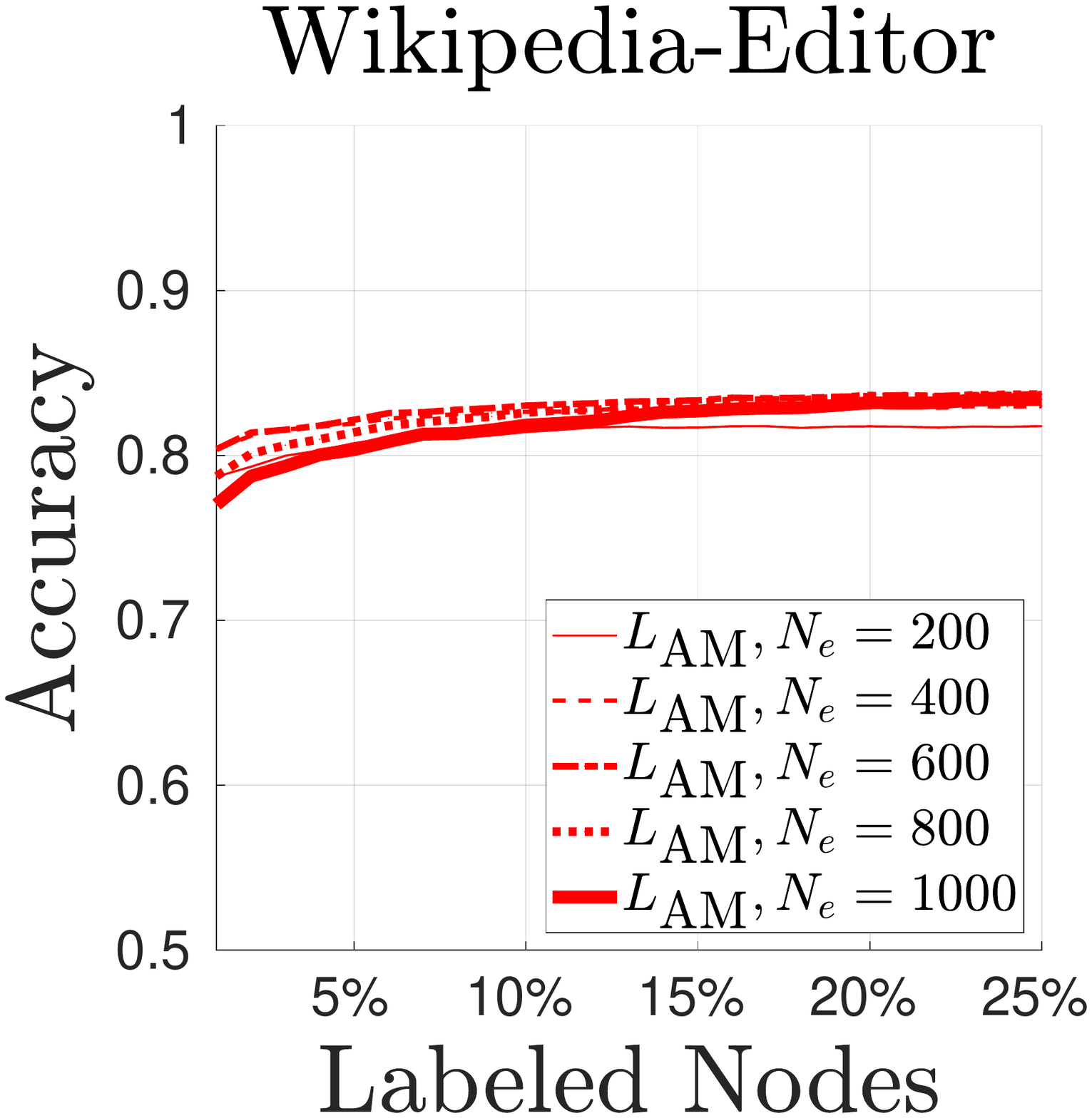}\hfill
\vspace{-10pt}
 \caption{Average classification accuracy with different amounts of labeled nodes given a fixed number of eigenvectors.
 Each row presents classification accuracy of dataset Wikipedia-RfA, Wikipedia-Elec, and Wikipedia-Editor. 
 Each column presents classification accuracy of  \textbf{GL}($L_\sym^+$), \textbf{GL}($Q_\sym^-$), \textbf{GL}($L_{SN}$), and \textbf{GL}($L_{AM}$).
\vspace{-10pt}
 }
 \label{fig:fixEigenvectorsRunLabeledNodes_SmoothPotential_V2}
\end{figure}

\begin{figure}[t]
 \centering
 \includegraphics[width=0.30\textwidth,trim=120 45 170 40,clip]{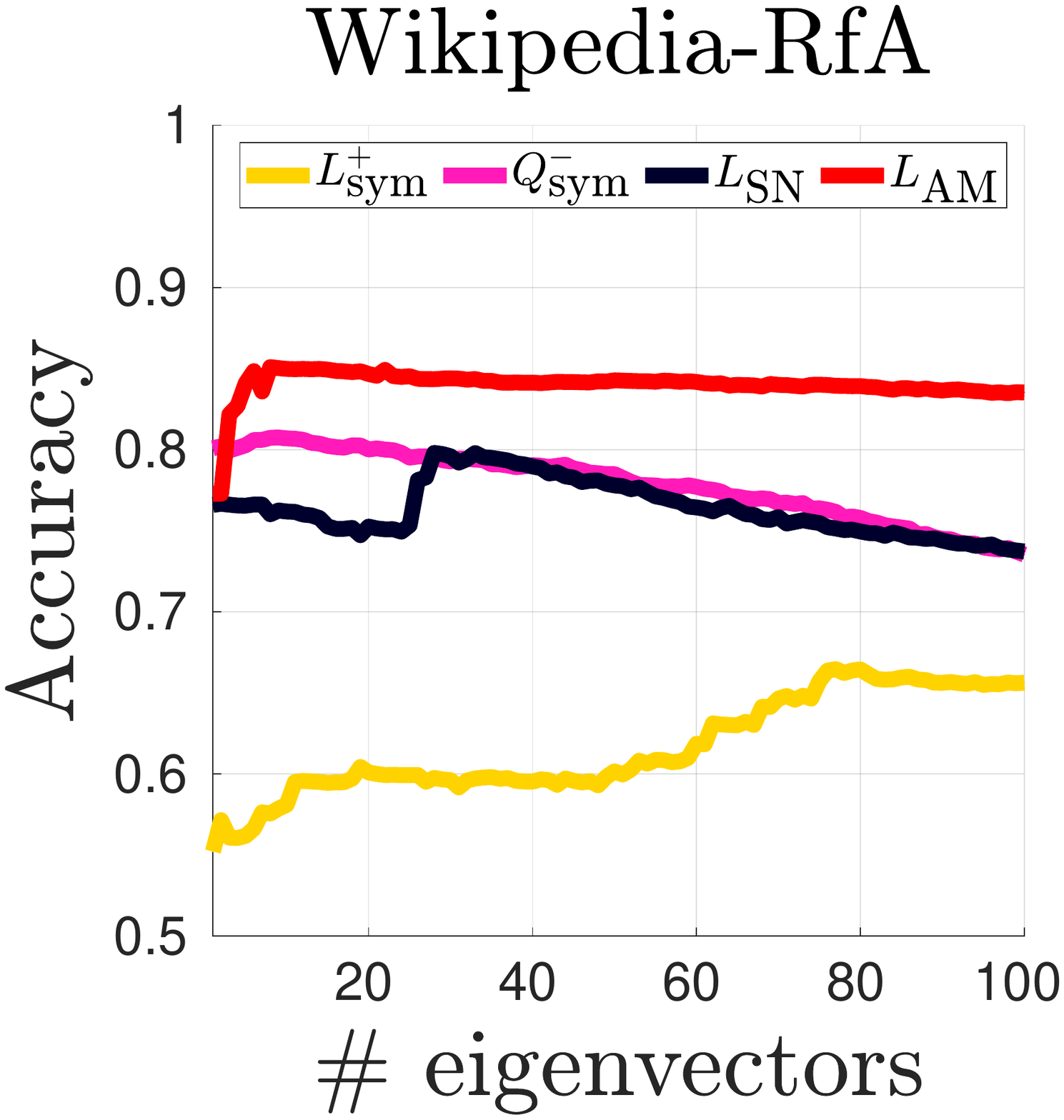}\hfill
 \includegraphics[width=0.30\textwidth,trim=120 45 170 40,clip]{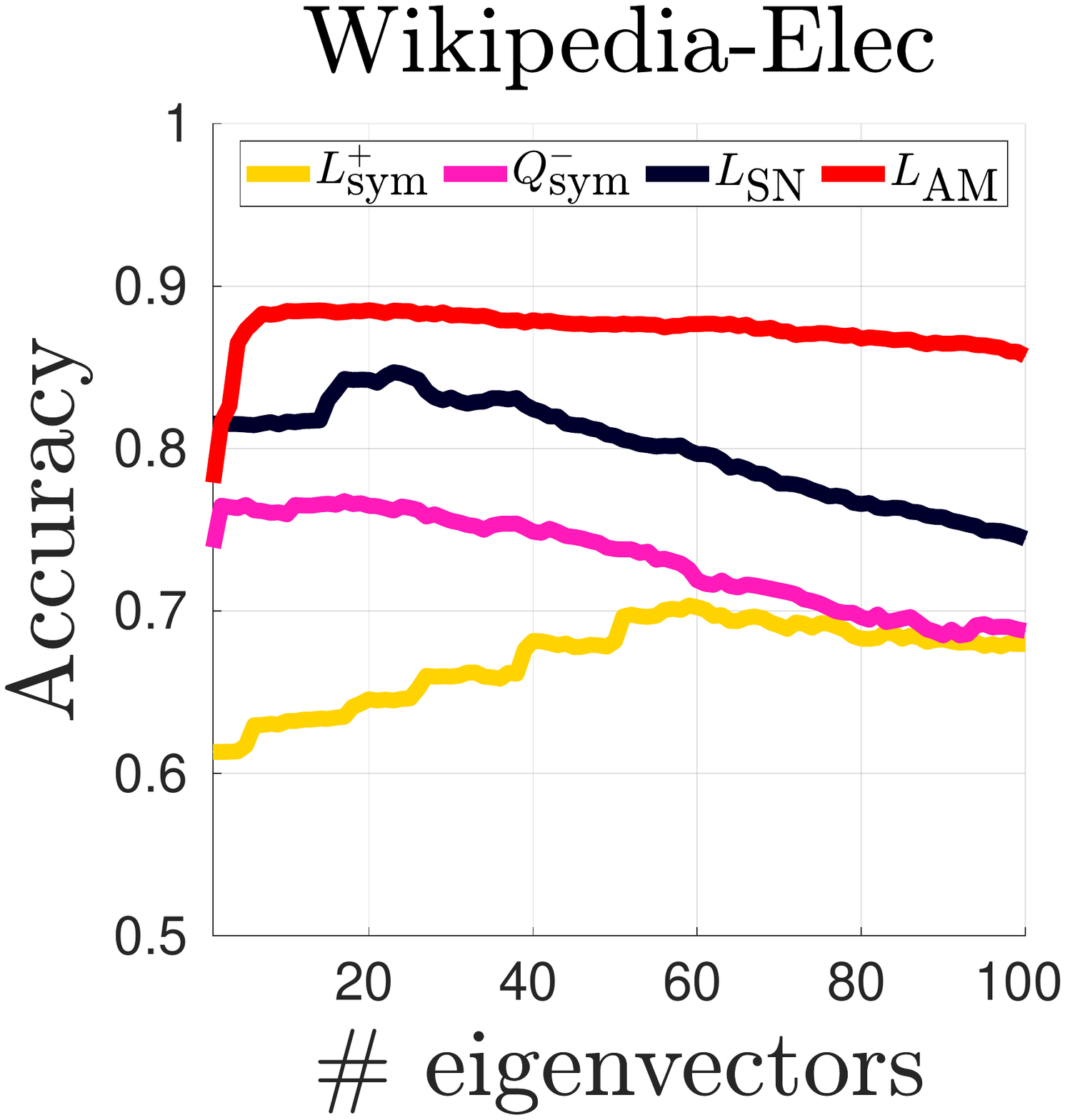}\hfill
 \includegraphics[width=0.30\textwidth,trim=120 45 170 40,clip]{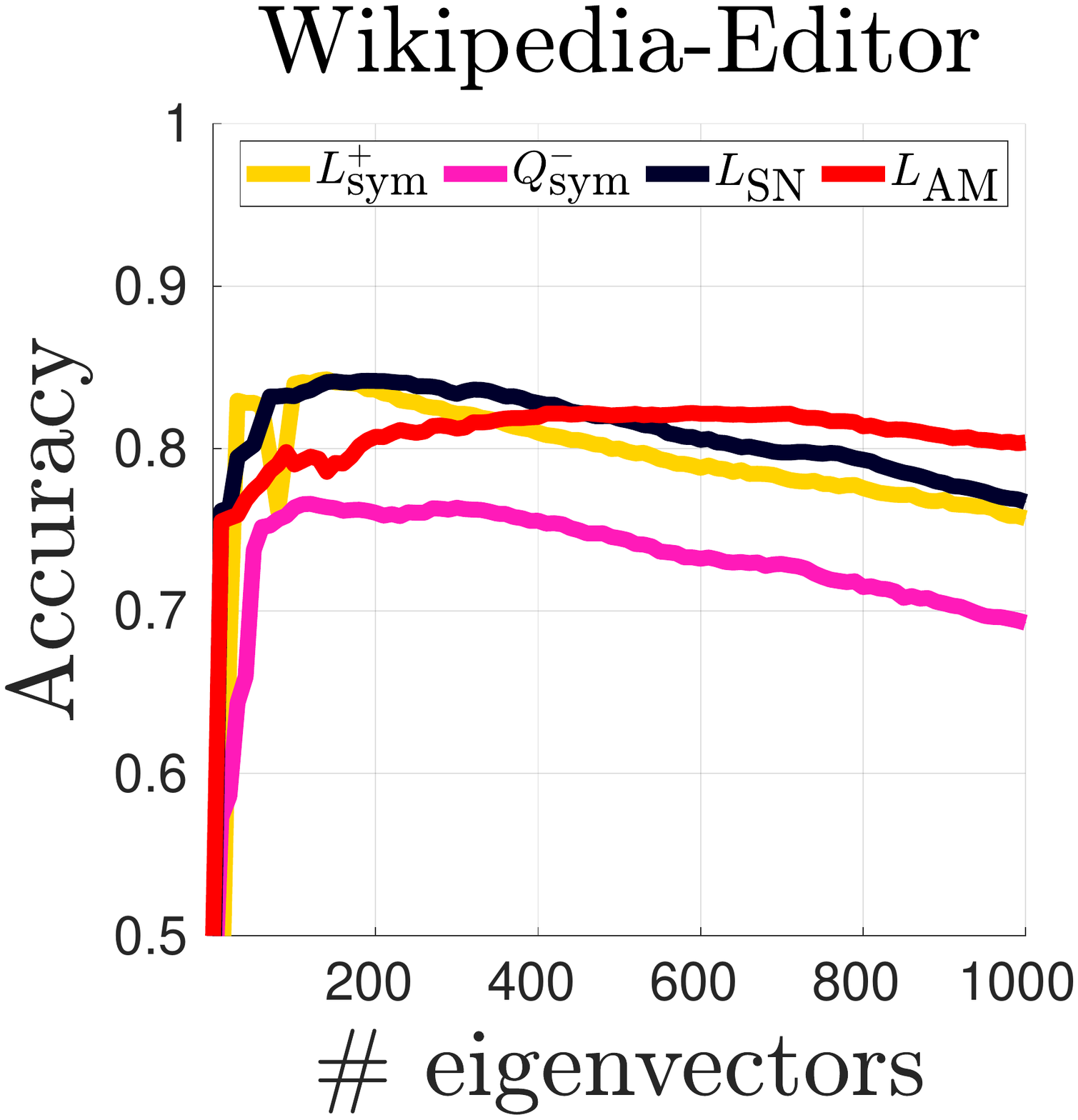}
 \caption{
 \pedro{
 Average classification accuracy with $5\%$ labeled nodes and different amounts of eigenvectors.
 Average accuracy is computed out of 10 runs. Our method based on Laplacians $L_{\textrm{SN}}$ and $L_{\textrm{AM}}$ consistently presents the best classification performance.
 }
 \vspace{-10pt}
 }
 \label{fig:V20-fig:fixLabeledNodesRunEigenvectors_SmoothPotential_V2}
\end{figure}

\subsection{\pedro{Effect of the Number of Labeled Nodes}}\label{subsubsec:EffectOfTheNumberOfLabeledNodes}
We now study how the classification accuracy of our method is affected by the amount of labeled nodes. 
For our method we fix the number of eigenvectors to $N_e\in\{20,40,60,80,100\}$ for Wikipedia-RfA and Wikipedia-Elec, and $N_e\!\in\!\{200,400,600,800,1000\}$ for Wikipedia-Editor. Given $N_e$, we evaluate our method
with different proportions of labeled nodes, going from $1\%$ to $25\%$ of the number of nodes $\abs{V}$.
%
%
%

The corresponding average classification accuracy is shown in Fig.~\ref{fig:fixEigenvectorsRunLabeledNodes_SmoothPotential_V2}.
As expected, we can observe that the classification accuracy increases with larger amounts of labeled nodes. Further, we can observe that this effect is more pronounced when larger amounts of eigenvectors $N_e$ are taken, i.e. the smallest classification accuracy increment is observed when the number of eigenvectors $N_e$ is $20$ for Wikipedia-RfA and Wikipedia-Elec and $100$ eigenvectors for Wikipedia-Editor.
\pedro{
Further, we can observe that overall our method based on \textbf{GL}($L_{SN}$) and \textbf{GL}($L_{AM}$) performs best, suggesting that blending the information coming from both positive and negative edges is beneficial for the task of node classification.
}

While our method based on signed Laplacians \textbf{GL}($L_{SN}$) and \textbf{GL}($L_{AM}$) overall presents the best performance, we can observe that they present a slightly difference when it comes to its sensibility to the amount of labeled nodes.
In particular, we can observe how the increment on classification accuracy \textbf{GL}($L_{SN}$) is rather clear, whereas with \textbf{GL}($L_{AM}$) the increment is  smaller. Yet, \textbf{GL}($L_{AM}$) systematically presents a better classification accuracy when the amount of labeled nodes is limited.
%
%
%

%

\subsection{\pedro{Effect of the Number of Eigenvectors}}\label{subsubsec:EffectOfTheNumberOfEigenvectors}

%
%
%
%
%
%

We now study how the performance of our method is affected by the number of eigenvectors given through different Laplacians. 
%
We fix the amount of labeled nodes to $5\%$ and consider different amounts of given eigenvectors. For datasets Wikipedia-RfA and Wikipedia-Elec we set the number of given eigenvectors $N_e$ in the range $N_e=1,\ldots,100$ and for Wikipedia-Editor in the range $N_e=1,10,\ldots,1000$. 
%

The average classification accuracy is shown in Fig.~\ref{fig:V20-fig:fixLabeledNodesRunEigenvectors_SmoothPotential_V2}.
For Wikipedia-RfA and Wikipedia-Elec we can see that the classification accuracy of \pedro{our method based on } \textbf{GL}($Q^-_\sym$) outperforms \pedro{our method based on the Laplacian} \textbf{GL}($L^+_\sym$) by a meaningful margin, suggesting that for \old{this}\pedro{the task of node classification} negative edges are more informative than positive edges.
Further, we can see that \textbf{GL}($L_{\text{AM}}$) consistently shows the highest classification accuracy \pedro{indicating that taking into account the information coming from both positive and negative edges is beneficial for classification performance. }

\pedro{For the case of Wikipedia-Editor the previous distinctions are not clear anymore. For instance, we can see that the performance of our method based on the Laplacian \textbf{GL}($L^+_\sym$) outperforms the case with  \textbf{GL}($Q^-_\sym$). Moreover, the information coming from positive edges presents a more prominent performance, being competitive to our method based on the Laplacian \textbf{GL}($L_{\text{SN}}$) when the number of eigenvectors is relatively small, whereas the case with the arithmetic mean Laplacian \textbf{GL}($L_{\text{AM}}$) presents a larger classification accuracy for larger amounts of eigenvectors.
}
%
Finally, we can see that in general our method first presents an improvement in classification accuracy, reaches a maximum and then decreases with the amount of given eigenvectors.

%


\subsection{\pedro{Joint Effect of the Number of Eigenvectors and Labeled Nodes}}\label{subsubsec:EffectoOfNumberOfEigenvectorsAndAmountOfLabeledNodes}
%
%
%
%
%
%
%

We now study the joint effect of the number of eigenvectors and the amount of labeled nodes in the classification performance of our method based on \textbf{GL}($L_{\text{SN}}$).
%
We let the number of eigenvectors $N_e\in\{10,20,\ldots,100\}$ for datasets Wikipedia-RfA and Wikipedia-Elec and $N_e\in\{100,200,\ldots,1000\}$ for dataset Wikipedia-Editor. Further, we let the amount of labeled nodes to go from $1\%$ to $25\%$.
%
The corresponding results are shown in Fig.~\ref{fig:numEigsVsLabeledNodes}, where we confirm that the classification accuracy consistently increases with larger amounts of labeled nodes. 
Finally, we can notice that the classification accuracy first increases with the amount of eigenvectors, it reaches a maximum, and then slightly decreases.
%
To better appreciate the performance of our method under various settings, we present the difference between the lowest and largest average classification accuracy in the bottom table of Fig.~\ref{fig:numEigsVsLabeledNodes}. 
We can see that the increments go from $25.13\%$ to $36.54\%$.
%
\setcounter{figure}{2}   
\begin{figure}[t]
    \centering

  \begin{subfigure}{\columnwidth}
  \centering
  \includegraphics[width=0.30\textwidth,trim=100 60 100 60]{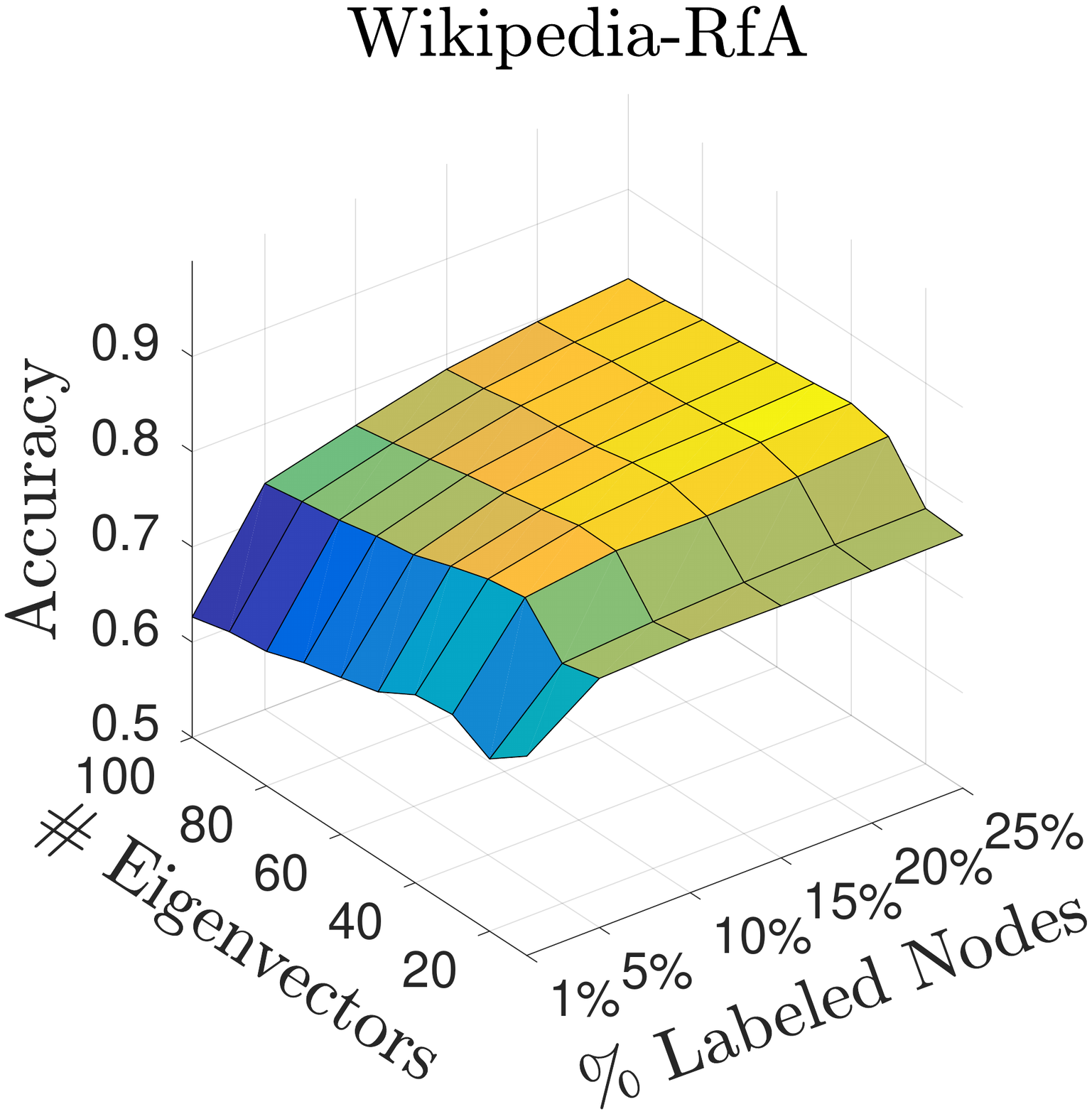}\hfill
  \includegraphics[width=0.30\textwidth,trim=100 60 100 60]{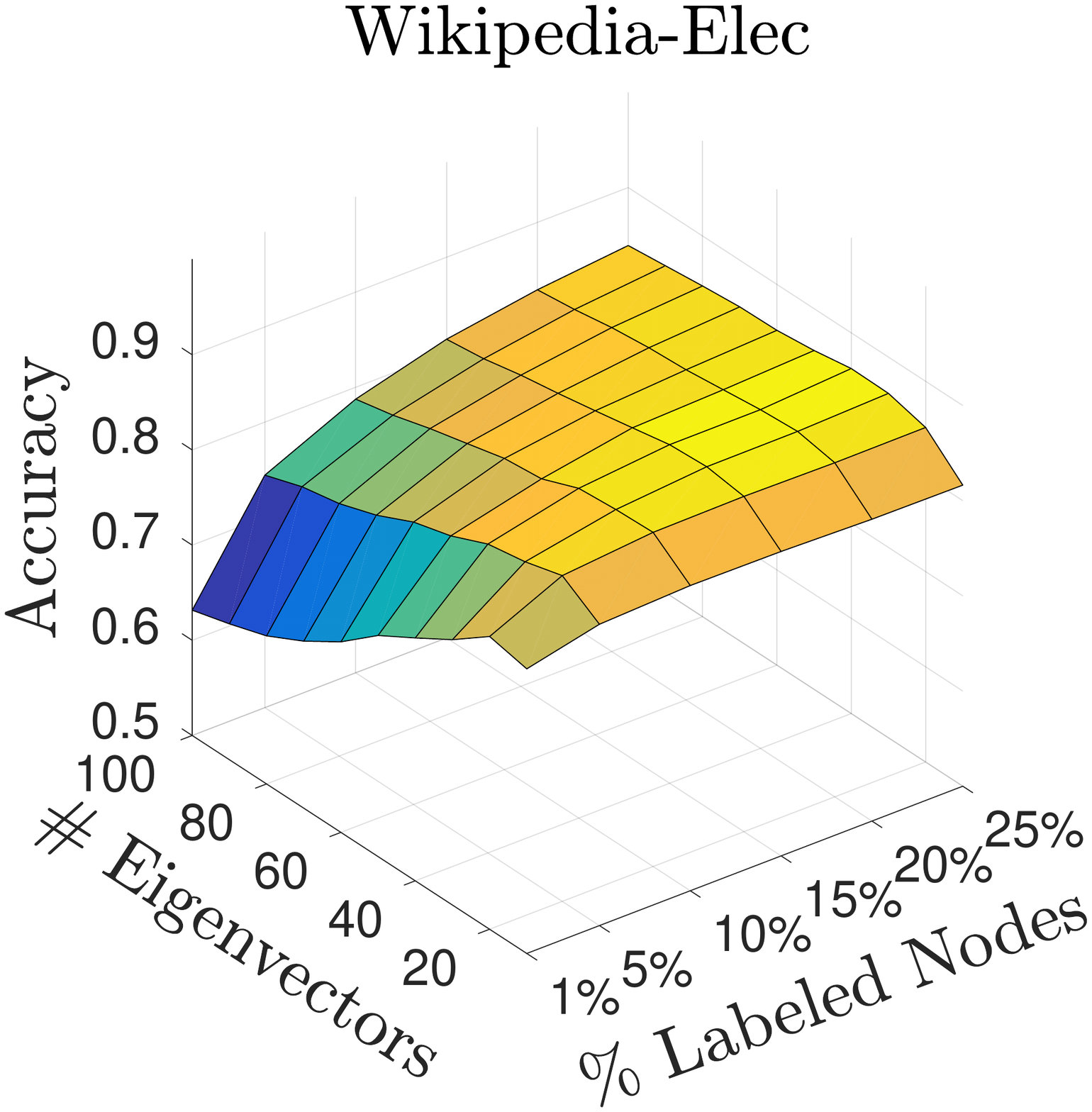}\hfill
  \includegraphics[width=0.30\textwidth,trim=100 60 100 60]{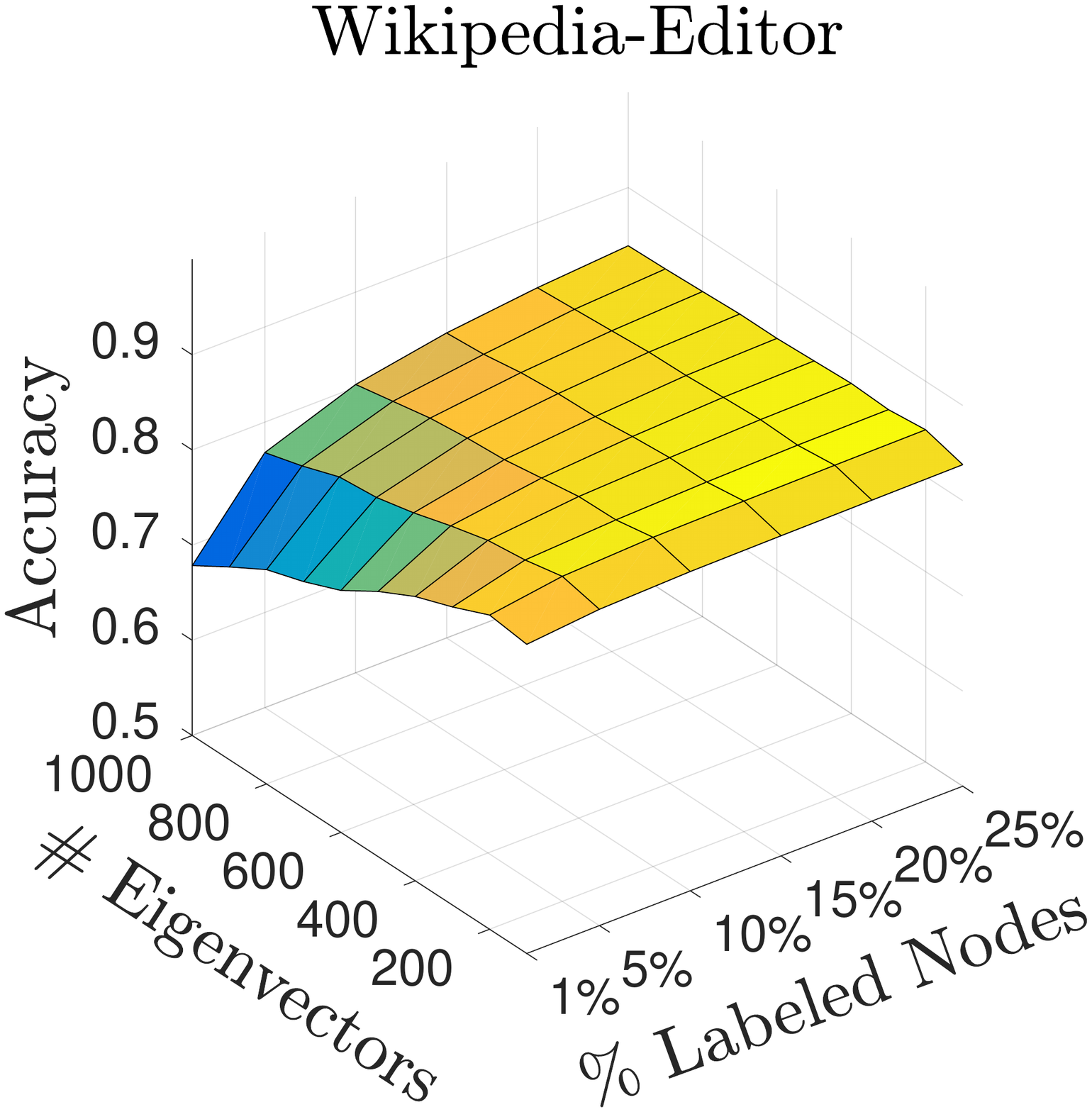}\hfill
  \vspace{10pt}
  \label{fig:numEigsVsLabeledNodes}
  \end{subfigure}
    \bigskip
    \begin{subfigure}{1\columnwidth}
        \centering
        \renewcommand\tabularxcolumn[1]{m{#1}}
        \renewcommand\arraystretch{1.3}
        \setlength\tabcolsep{2pt}
    \begin{tabularx}{\linewidth}{*{4}{>{\centering\arraybackslash}X}}
    \hline
Dataset          & Lowest Accuracy & Largest Accuracy & Increment \\ \hline
wikipedia-Elec   & 0.6317          & 0.8625           & 36.54$\%$ \\ \hline
wikipedia-RfA    & 0.6264          & 0.8280           & 32.17$\%$ \\ \hline
wikipedia-Editor & 0.6785          & 0.8491           & 25.13$\%$ \\ \hline
    \hline
    \end{tabularx}
    \end{subfigure}
    \vspace{-10pt}
\caption{Top: Average classification accuracy of our method with \textbf{GL}($L_{SN}$) under different number of eigenvectors and different amounts of labeled nodes. 
Bottom:
Lowest and largest average classification accuracy of \textbf{GL}($L_{\text{SN}}$) per dataset.
}
    \label{fig:numEigsVsLabeledNodes}
\end{figure}
\setcounter{figure}{3}   
\begin{figure}[t]
 \begin{subfigure}[b]{0.22\textwidth}
 \includegraphics[width=\textwidth,trim=160 0 180 40,clip]{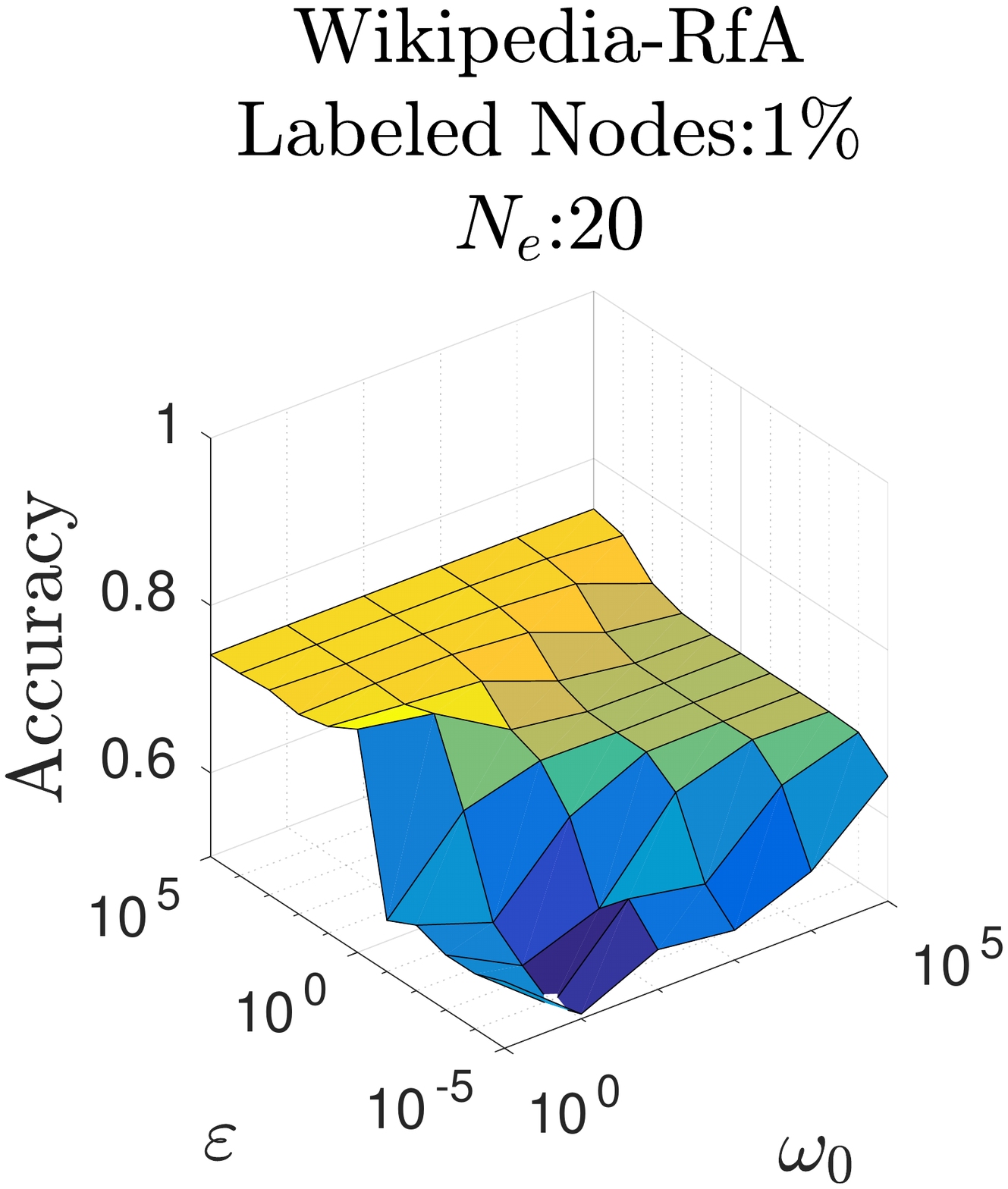}
  \vspace*{-8mm}
 \caption{}
 \label{fig:fidelityVSinterface-WikipediaRfA}
 \end{subfigure}%
 \hfill %
 \begin{subfigure}[b]{0.22\textwidth}
 \includegraphics[width=\textwidth,trim=160 0 180 40,clip]{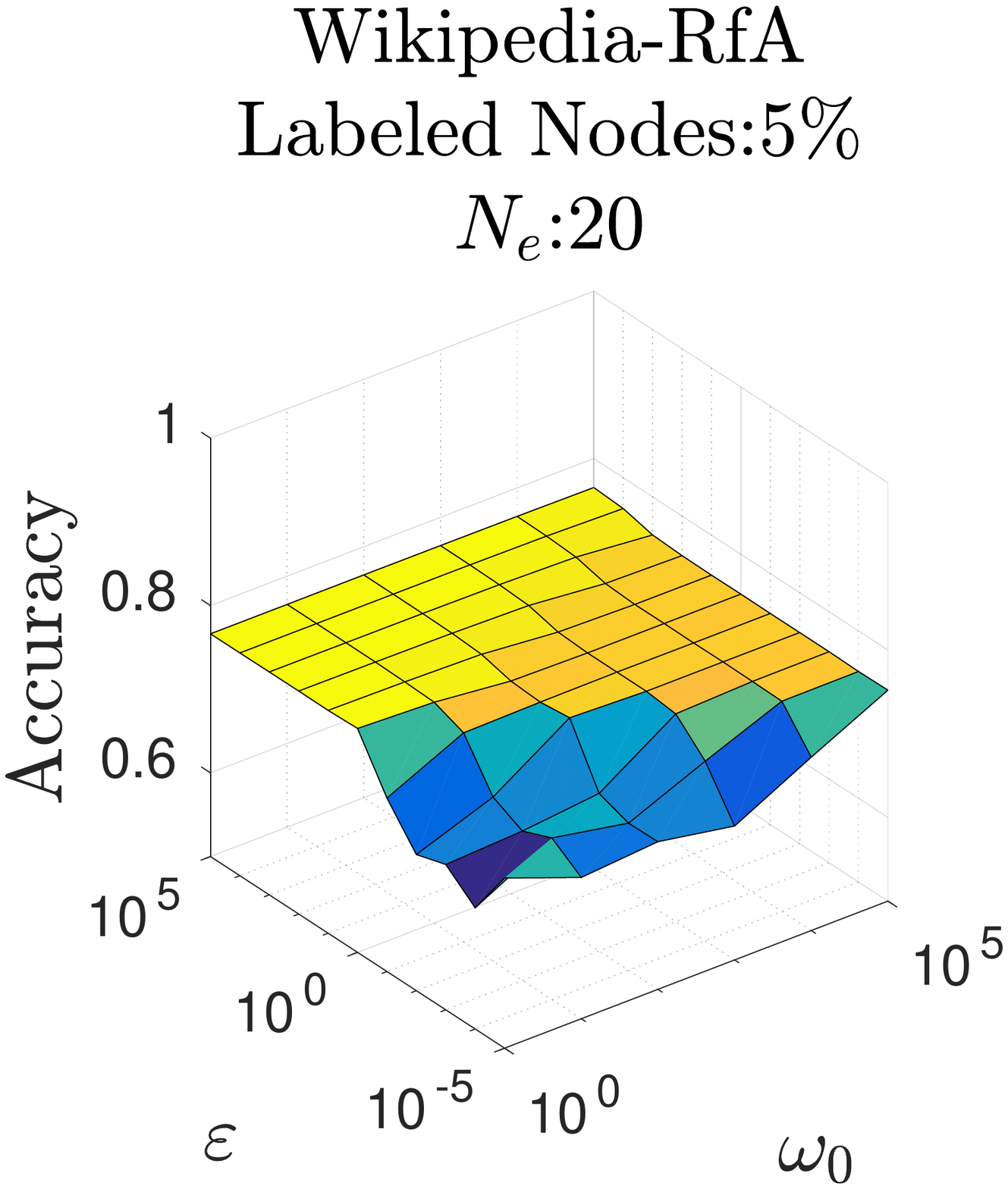}
  \vspace*{-8mm}
 \caption{}
 \end{subfigure}%
 \hfill %
 \begin{subfigure}[b]{0.22\textwidth}
 \includegraphics[width=\textwidth,trim=160 0 180 40,clip]{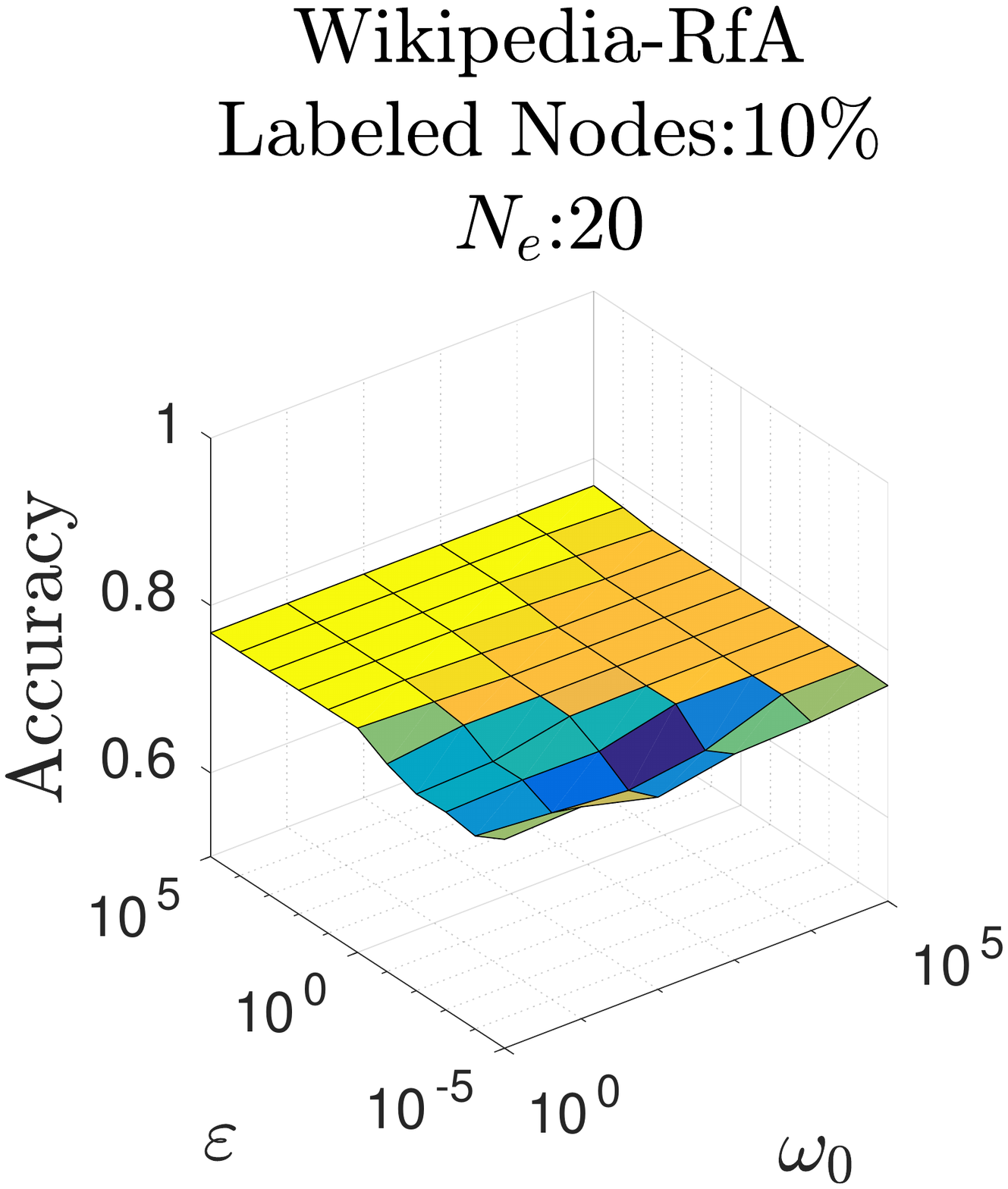}
  \vspace*{-8mm}
 \caption{}
 \end{subfigure}%
 \hfill %
 \begin{subfigure}[b]{0.22\textwidth}
 \includegraphics[width=\textwidth,trim=160 0 180 40,clip]{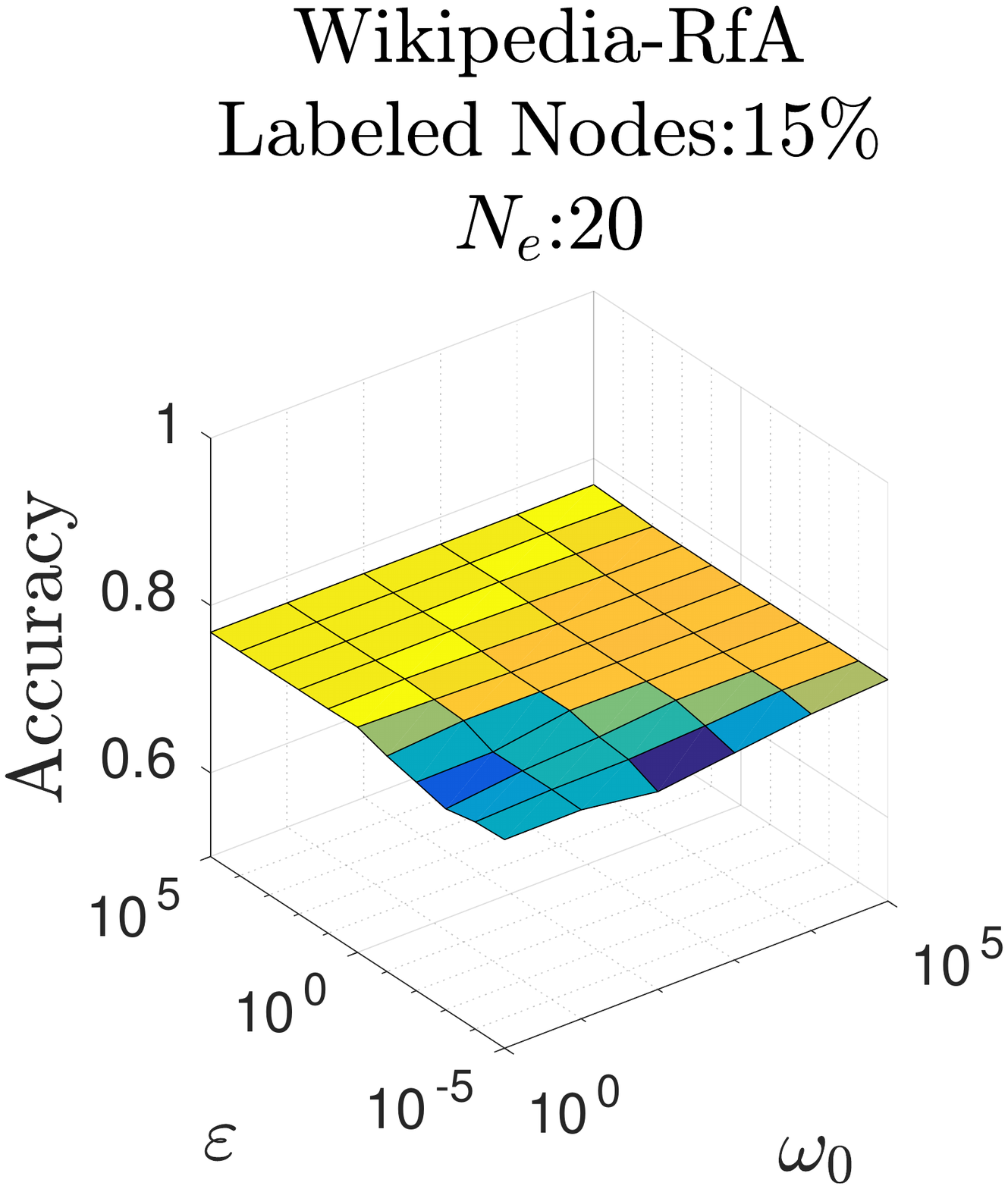}
  \vspace*{-8mm}
 \caption{}
 \end{subfigure}%
 \hfill %
 \\
 \begin{subfigure}[b]{0.22\textwidth}
 \includegraphics[width=\textwidth,trim=160 0 180 40,clip]{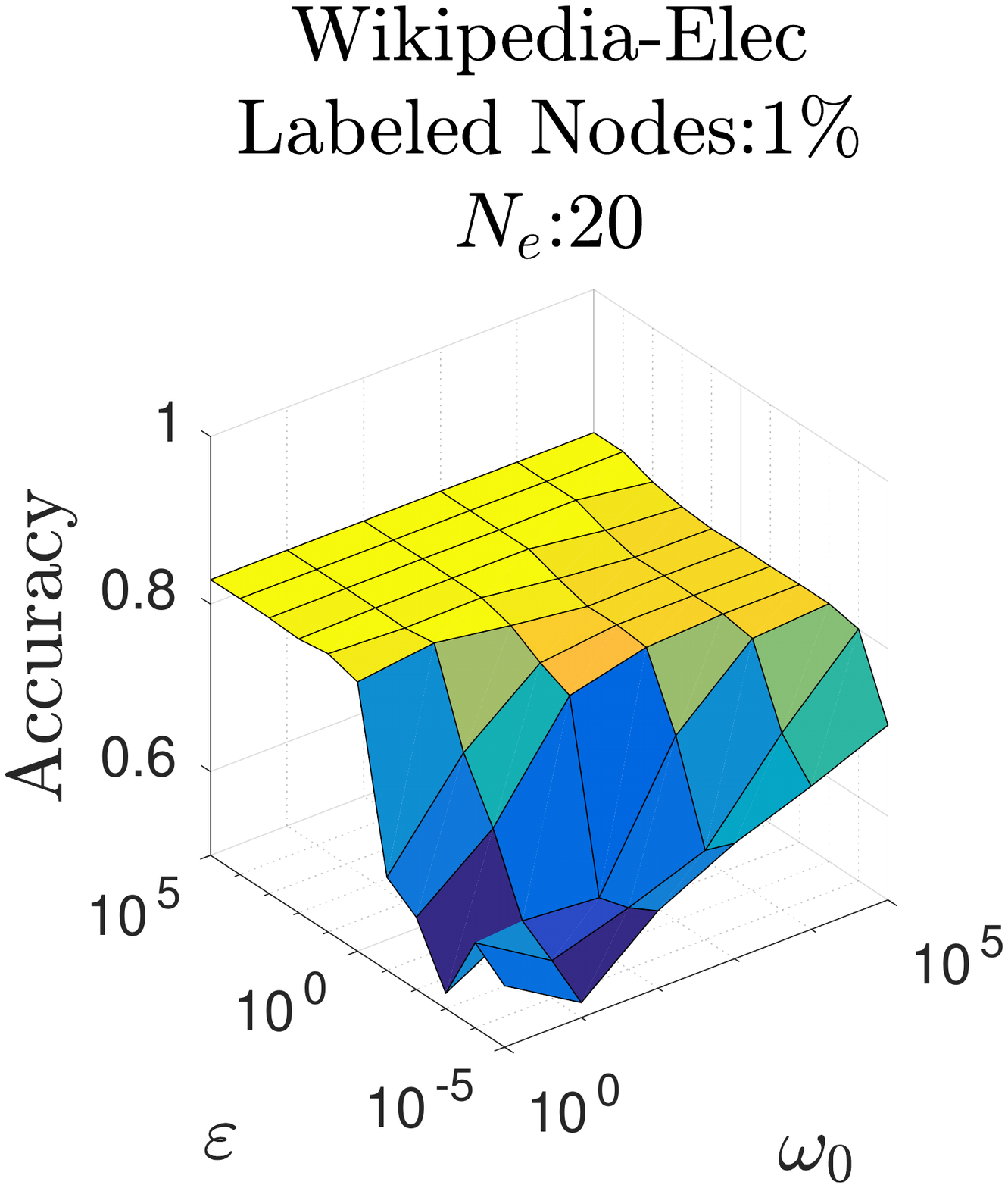}
 \vspace*{-8mm}
 \caption{}
 \label{fig:fidelityVSinterface-WikipediaElec}
 \end{subfigure}%
 \hfill %
 \begin{subfigure}[b]{0.22\textwidth}
 \includegraphics[width=\textwidth,trim=160 0 180 40,clip]{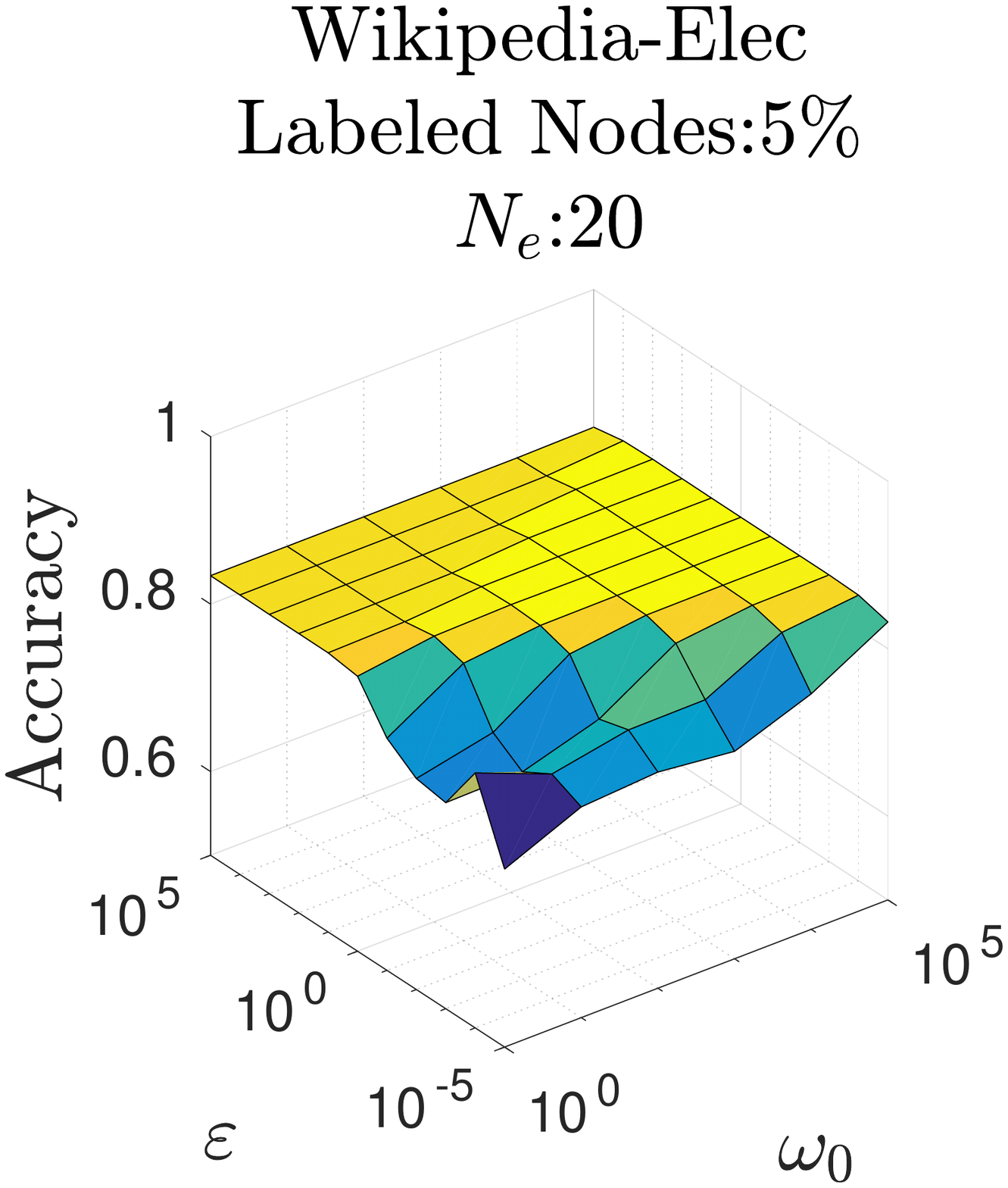}
 \vspace*{-8mm}
 \caption{}
 \end{subfigure}%
 \hfill %
 \begin{subfigure}[b]{0.22\textwidth}
 \includegraphics[width=\textwidth,trim=160 0 180 40,clip]{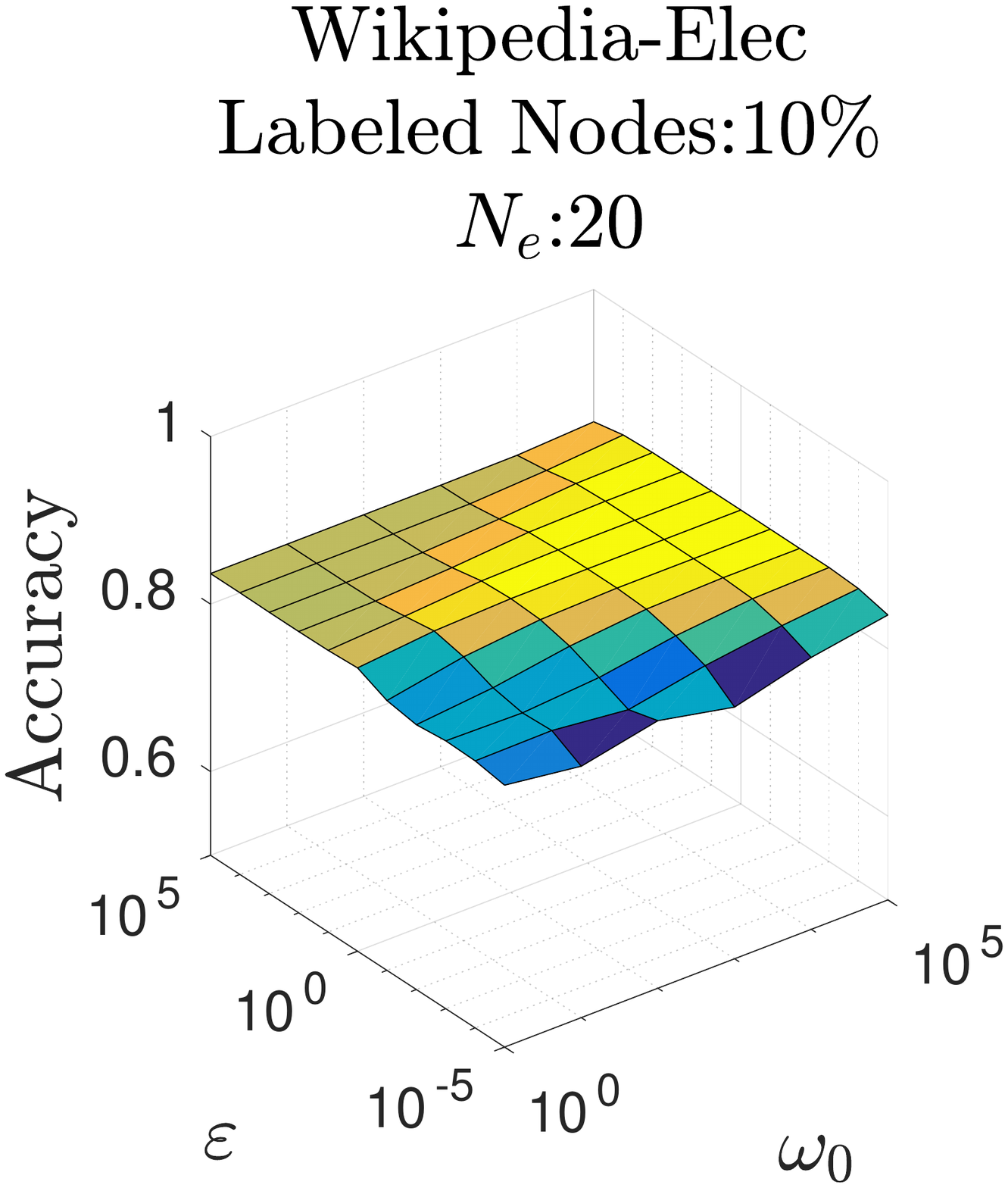}
 \vspace*{-8mm}
 \caption{}
 \end{subfigure}%
 \hfill %
 \begin{subfigure}[b]{0.22\textwidth}
 \includegraphics[width=\textwidth,trim=160 0 180 40,clip]{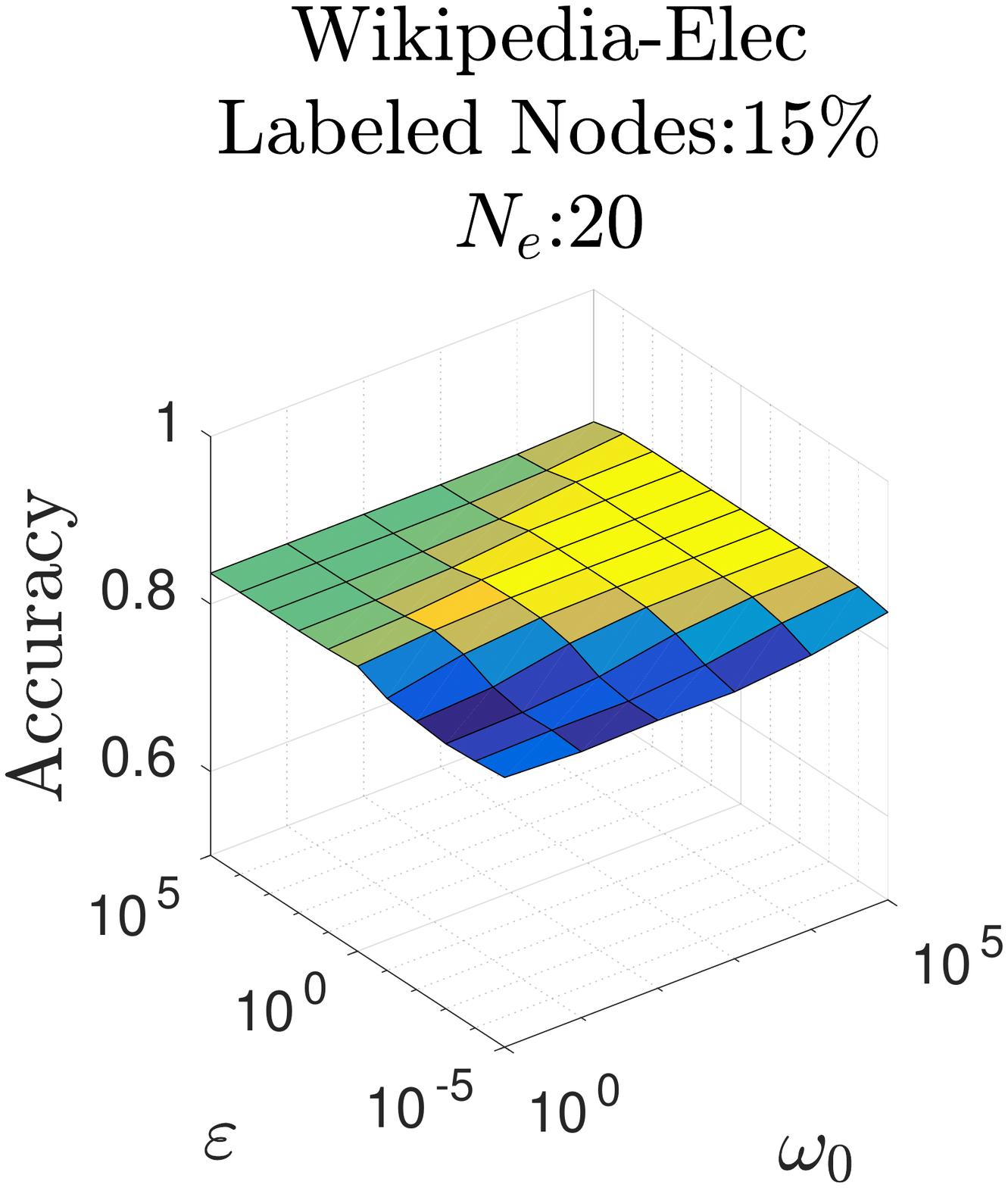}
 \vspace*{-8mm}
 \caption{}
 \end{subfigure}%
 \hfill
\\
  \begin{subfigure}[b]{0.22\textwidth}
 \includegraphics[width=\textwidth,trim=160 0 180 40,clip]{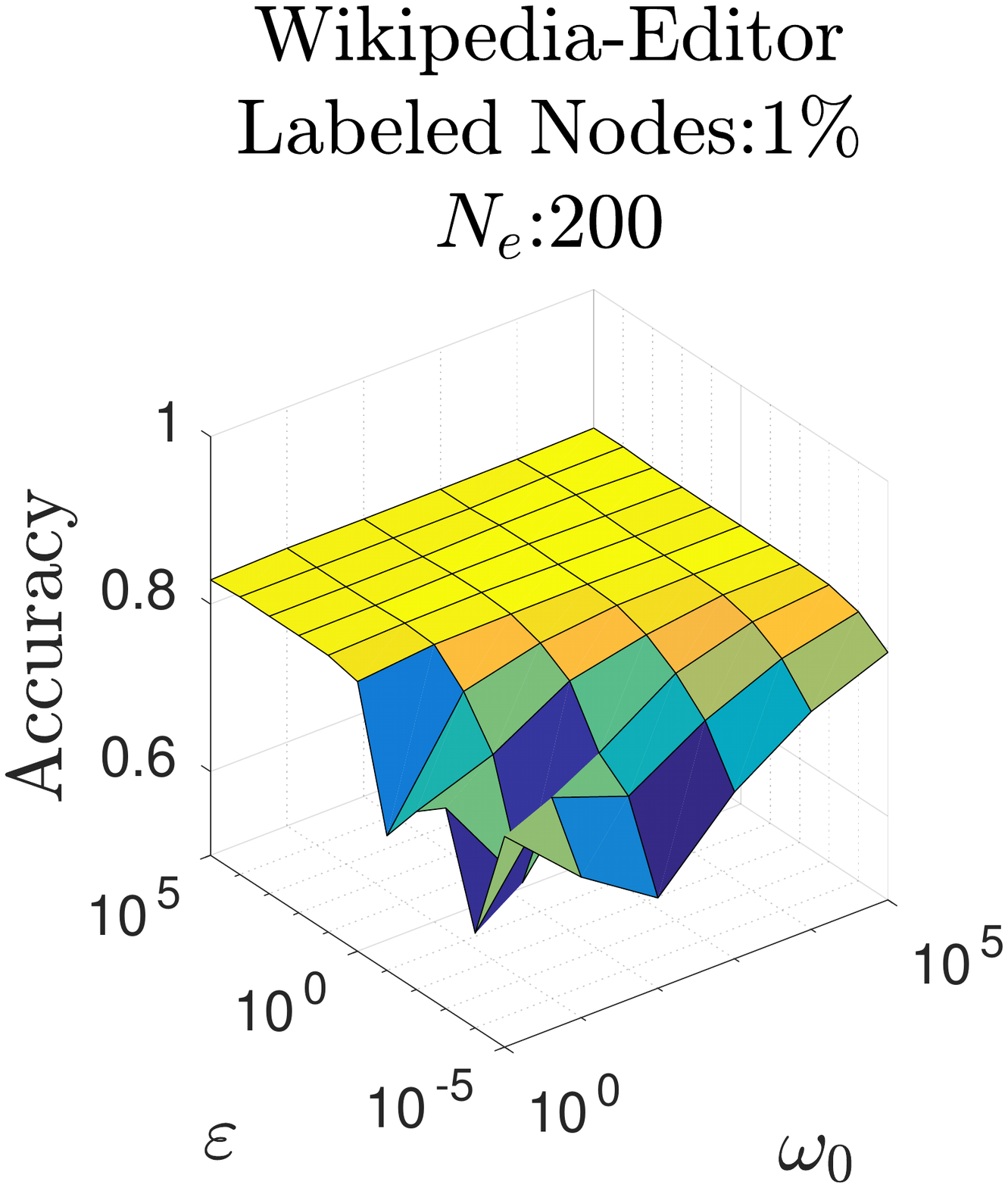}
  \vspace*{-8mm}
 \caption{}
 \label{fig:fidelityVSinterface-WikipediaEditor}
 \end{subfigure}%
 \hfill %
 \begin{subfigure}[b]{0.22\textwidth}
 \includegraphics[width=\textwidth,trim=160 0 180 40,clip]{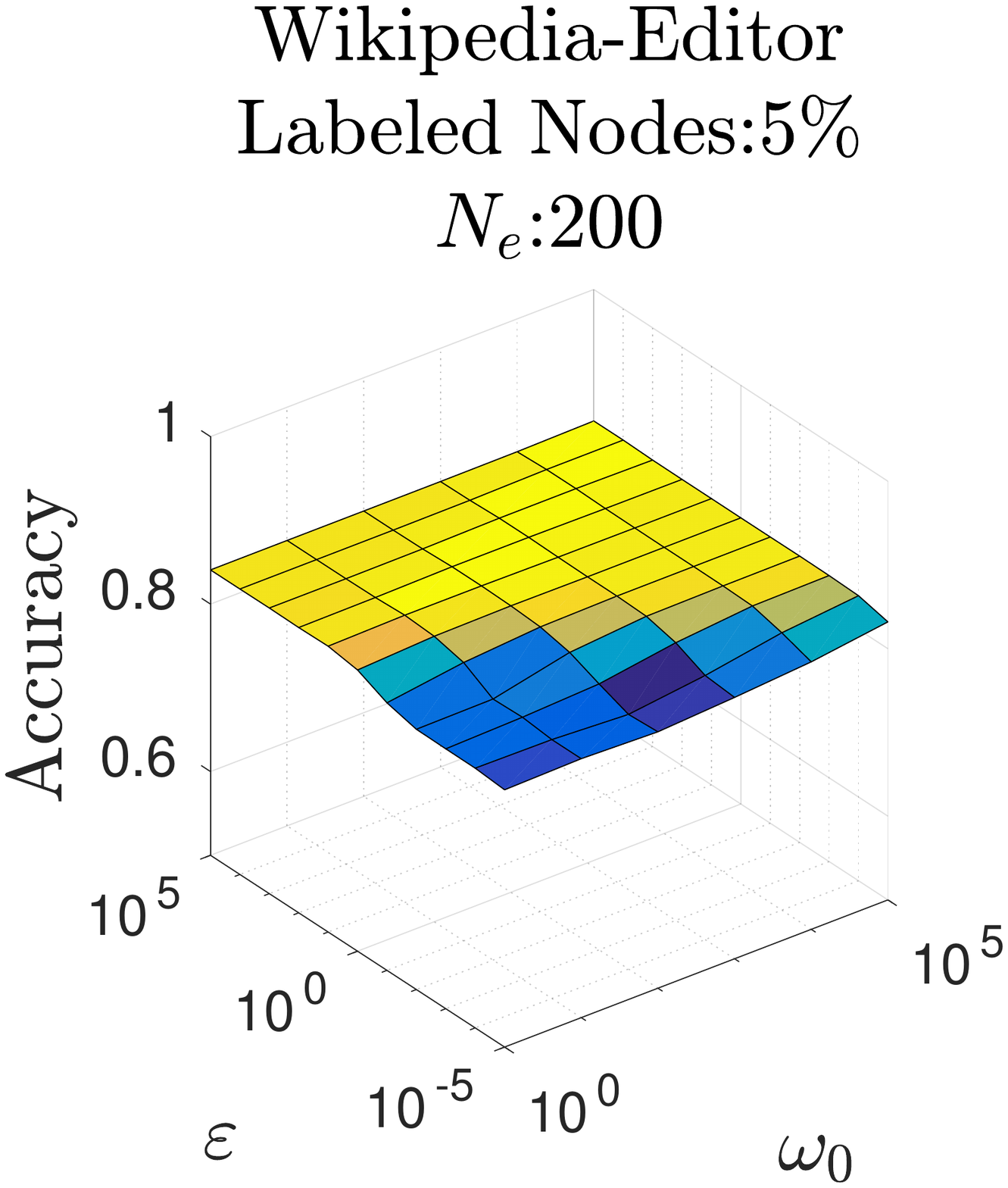}
  \vspace*{-8mm}
 \caption{}
 \end{subfigure}%
 \hfill %
 \begin{subfigure}[b]{0.22\textwidth}
 \includegraphics[width=\textwidth,trim=160 0 180 40,clip]{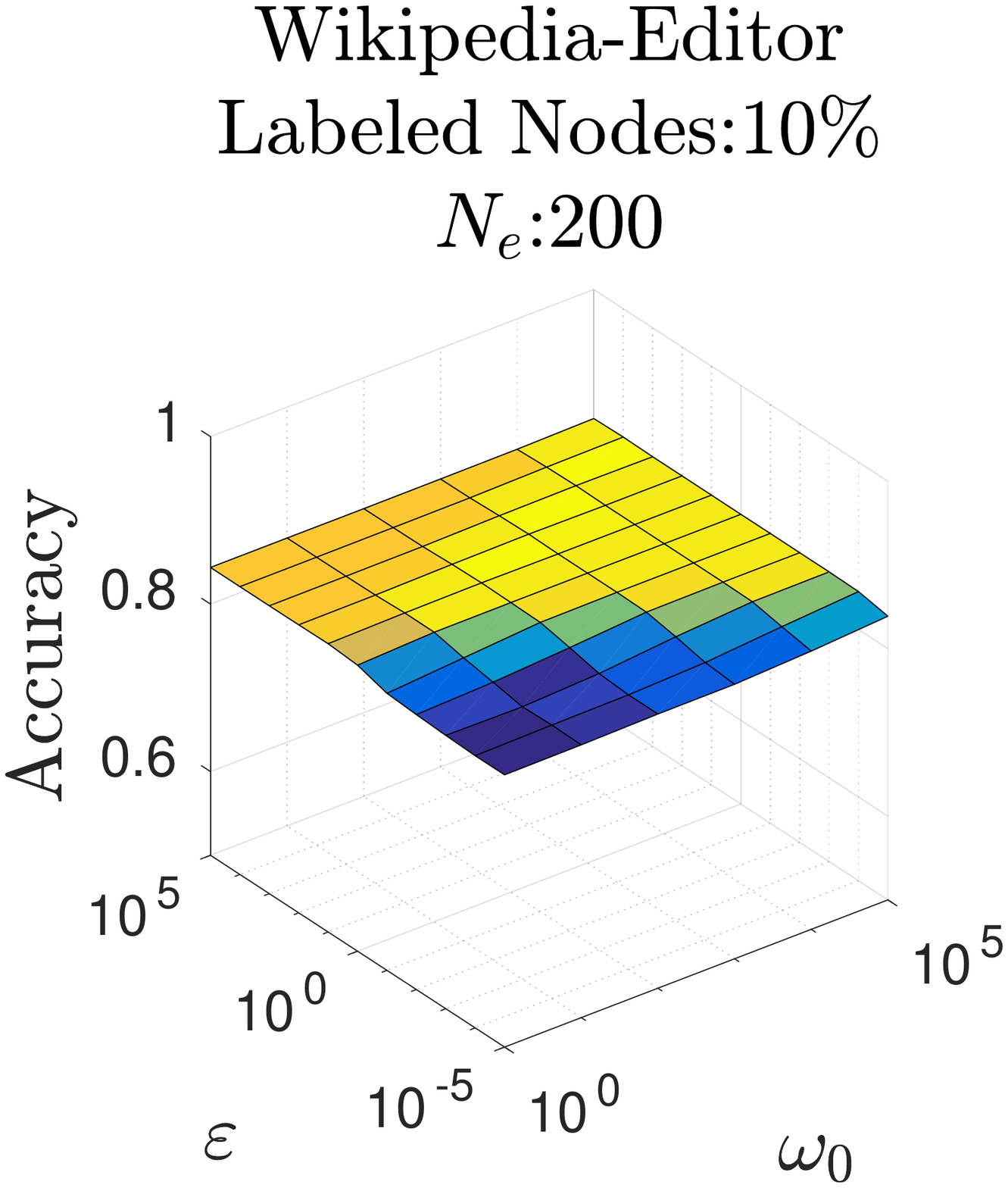}
  \vspace*{-8mm}
 \caption{}
 \end{subfigure}%
 \hfill %
 \begin{subfigure}[b]{0.22\textwidth}
 \includegraphics[width=\textwidth,trim=160 0 180 40,clip]{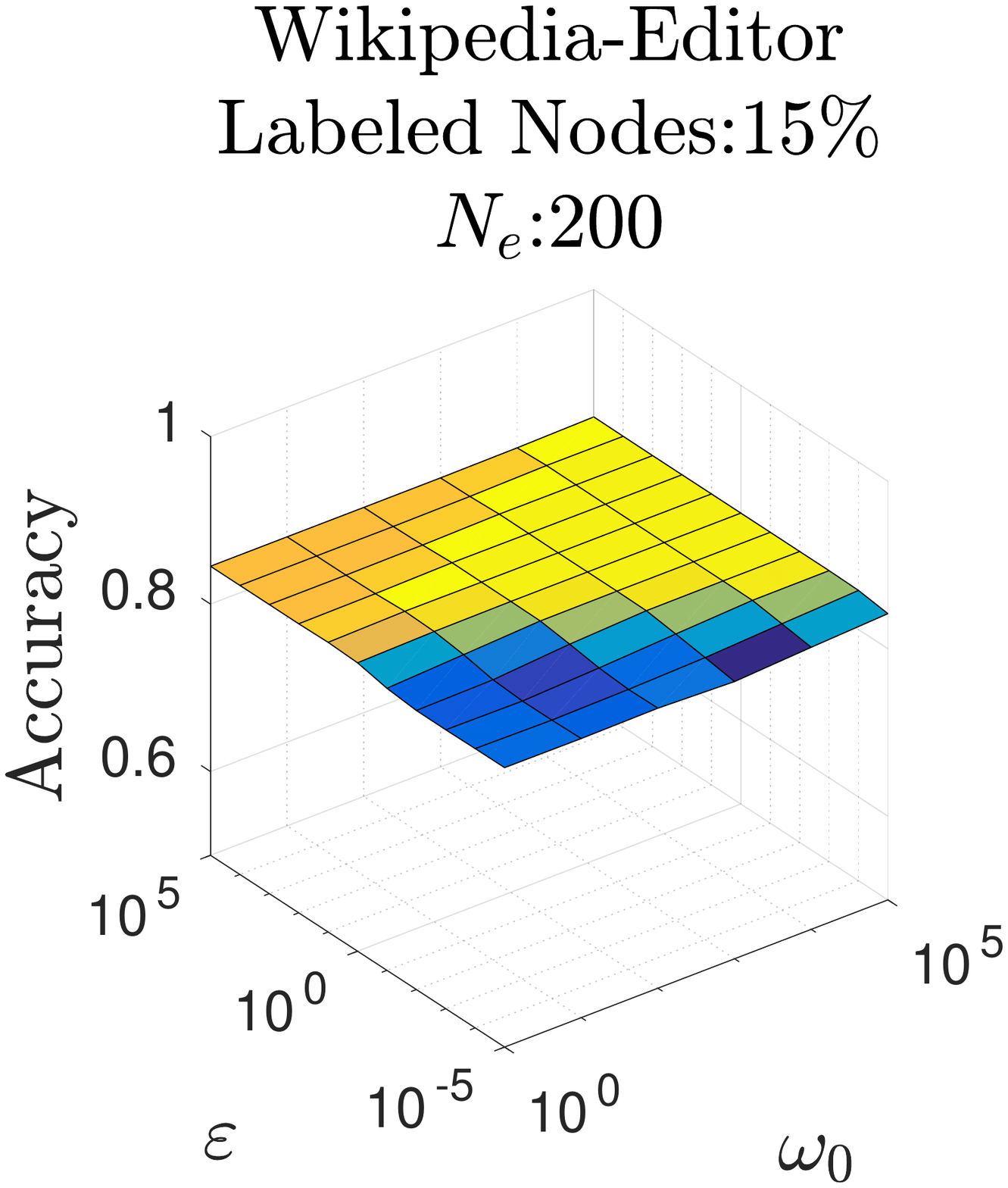}
  \vspace*{-8mm}
 \caption{}
 \end{subfigure}%
 \hfill %
  \caption{
  Average classification accuracy of our method based on \textbf{GL}($L_{SN}$) with different values of fidelity ($\omega_0$) and interface ($\varepsilon$).
  Columns (from left to right): amount of labeled nodes: $1\%,5\%,10\%,15\%$.
 Rows (from top to bottom): classification accuracy on datasets Wikipedia-RfA, Wikipedia-Elec, and Wikipedia-Editor. 
 }
 \label{fig:fidelityVSinterface}
\end{figure}
\subsection{Joint effect of fidelity ($\omega_0$) and interface ($\varepsilon$) parameters}

\label{subsubsec:parameterAnalysis}

We now study the effect of fidelity ($\omega_0$) and interface ($\varepsilon$) parameters on the classification accuracy of our method based on \textbf{GL}($L_{SN}$).
%
We fix the number of eigenvectors to $N_e=20$, and let the amount of labeled nodes to go from $1\%$ to $15\%$. Further, we set the fidelity parameter $\omega_0$ to take values in $\{10^0,10^1,\ldots,10^5\}$ and the interface parameter $\varepsilon$ to take values in $\{10^{-5},10^{-4},\ldots,10^4,10^5\}$.
%
%
The results are shown in Fig.~\ref{fig:fidelityVSinterface}.
%
We present the following observations:

%
\pedro{
First: we can see that the larger the amount of labeled nodes, the smaller is the effect of parameters $(\omega_{0}, \varepsilon)$. In particular, we can observe that when the amount of labeled nodes is at least $10\%$ of the number of nodes, then the parameter effect of $(\omega_{0}, \varepsilon)$ is small, in the sense that the classification accuracy remains high.
}

\pedro{
Second: we can see that there is a relationship between the fidelity parameter $\omega_{0}$ and the interface parameter $\varepsilon$ describing a \textit{safe region}, in the sense that the classification accuracy is not strongly affected by the lack of large amounts of labeled nodes. In particular, we can observe that this region corresponds to the cases where the interface parameter $\varepsilon$ is larger than the fidelity parameter $\omega_{0}$, i.e. 
$\varepsilon(k_1) > \omega_0(k_2)$
where $\varepsilon(k_1)=10^{k_1}$ and $\omega_0(k_2)=10^{k_2}$, with $k_1\in\{10^0,10^1,\ldots,10^5\}$
and $k_2\in\{10^{-5},10^{-4},\ldots,10^4,10^5\}$.
This can be well observed through a slightly triangular region particularly present for the case where the amount of labeled nodes is $1\%$ on 
all
datasets,
which is depicted in Figs.~\ref{fig:fidelityVSinterface-WikipediaRfA},
\ref{fig:fidelityVSinterface-WikipediaElec}, 
and
\ref{fig:fidelityVSinterface-WikipediaEditor}
.
}

\section{Conclusion}

We have illustrated that the semi-supervised task of node classification in signed networks can be performed via a natural extension of diffuse interface methods by taking into account suitable signed graph Laplacians.
We have shown that different signed Laplacians provide different classification performances under real world signed networks. In particular, we have observed that negative edges provide a relevant amount of information, leading to an improvement in classification performance when compared to the unsigned case.
As future work the task of non-smooth potentials can be considered, together with more diverse functions of matrices that would yield different kinds of information merging of both positive and negative edges.

\bibliographystyle{splncs04}

\newpage\clearpage
%

\section{Vector-valued formulation}\label{appendix:multiclass}\label{subsec:vectorGH}

This section contains further details of multi-class case of the approach proposed in Section~\ref{section:DiffuseInterfaceMethods}.

Garcia-Cardona et al.~\cite{GarMBFP14} as well as Merkurjev et al.~\cite{MerGBFP14} have extended 
the use of the diffuse interface model based on the generalized Ginzburg--Landau energy
to multi-class segmentation of high-dimensional data on graphs. For hypergraphs this was done in \cite{BosKS16}.
We introduce the matrix 
$U=({u}_{1},\ldots,{u}_{n})^{T}\in\mathbb{R}^{n\times K}$, where the $m$th component of the vector ${u}_{i}\in\mathbb{R}^{K}$ indicates the strength for data point or graph vertex $i$ to belong to class $m$. Interpreting this is the sense of a probability distribution for vertex $i$ belonging to class $m,$ we need to make sure that the sum of the entries in one row will sum to one. For this we now for each node $i$ force the vector ${u}_{i}$ to be an element of the Gibbs simplex $\Sigma^{K}$
\begin{displaymath}
 \Sigma^{K}:=\left\{(x_{1},\ldots,x_{K})^{T}\in[0,1]^{K}\left\vert\, \sum_{l=1}^{K}{x_{l}}=1\right.\right\}.
\end{displaymath}
The vector-valued Ginzburg--Landau energy functional on graphs as in Section~~\ref{section:DiffuseInterfaceMethods} of the main paper generalizes to the multi-class case as
\begin{equation}
\label{gl_vv}
E(U)=\frac{\eps}{2} \textup{trace}(U^{T}\Smatrix  U)+
\frac{1}{2\eps}\sum_{i\in V}{\left(\prod_{l=1}^{K}{\frac{1}{4}||{u}_{i}-{\bf e}_{l}}||_{L_{1}}^{2}\right)}+
\sum_{i\in V}{\frac{\omega_i}{2}||\hat{u}_{i}-{u}_{i}||^{2}_{L_{2}}}.
\end{equation}
Let us explain this energy in more detail. We again have an energy term given by
$$\frac{\eps}{2} \textup{trace}(U^{T}\Smatrix  U),$$ which again induces smoothness and adds clustering information to the functional. The role played by this part mirrors the energy term
$\frac{\eps}{2} u^{T}\Smatrix u$ for the binary classification problem. The vector-valued potential 
$$
\frac{1}{2\eps}\sum_{i\in V}{\left(\prod_{l=1}^{K}{\frac{1}{4}||{u}_{i}-{\bf e}_{l}}||_{L_{1}}^{2}\right)}
$$
enforces that the components of $U$ are either $0$ or $1$.  The term that incorporates the already labeled information is here
$$
\sum_{i\in V}{\frac{\omega_i}{2}||\hat{u}_{i}-{u}_{i}||^{2}_{L_{2}}}
$$
with $\omega_i$ the penalty parameter analogous to the two-classes classification case and with $\hat U=(\hat{u}_{1},\ldots,\hat{u}_{n})^{T}$ representing the already labeled data. Here, 
${\bf e}_{l}\in\mathbb{R}^{K}$ being the vector whose $l$-th component equals one and all other components
vanish. Note that the vectors ${\bf e}_{1},\ldots,{\bf e}_{K}$ correspond to the perfect classification outcome.
The authors in \cite{GarMBFP14,MerGBFP14} use an $L_{1}$-norm for the potential term (the middle term in (\ref{gl_vv})) since it prevents an undesirable minimum from occurring at the center of the  simplex, as it would be the case with an $L_{2}$-norm for large $K$, so as to avoid an undesired and hence useless classification result.

The same convexity splitting scheme as in Section~\ref{section:DiffuseInterfaceMethods} of the main is used to minimize the Ginzburg--Landau functional in the phase-field approach. This results in
\begin{equation}
 \label{CS_vv}
\frac{U^{(t+1)}-U^{(t)}}{\tau}+\eps \Smatrix  U^{(t+1)}+cU^{(t+1)}=-\frac{1}{2\eps}T(U^{(t)})+cU^{(t)}+\omega(\hat{U}-U^{(t)}),
\end{equation}
where the elements $T_{ik}$ of the matrix $T(U^{(t)})$ are given as
\begin{displaymath}
 T_{ik}=\sum_{l=1}^{K}{\frac{1}{2}\left(1-2\delta_{kl}\right)||\bar{u}_{i}-{\bf e}_{l}||_{L_{1}}}\prod_{m=1,m\neq l}^{K}{\frac{1}{4}||\bar{u}_{i}-{\bf e}_{m}}||_{L_{1}}^{2},
\end{displaymath}
which represents the derivative of the potential term. Note that $\omega$ is a diagonal matrix containing the $\omega_1,\ldots\omega_K$. The parameter $c\geq\omega_0+\frac{1}{\varepsilon}$ arises from the convexity splitting. 
As before, we assume that $U$ is evaluated at the new time-point, 
whereas $U^{(t)}$ indicates the previous time-point. 
Using the eigendecomposition $\Smatrix =\Phi\Lambda\Phi^{T}$ and multiplying (\ref{CS_vv}) by $\Phi^{T}$ from the left, we obtain
\begin{equation}
 \label{CS_vv_2}
 \mathcal{U}^{(t+1)}=B^{-1}\left[(1+c\tau)\mathcal{U}^{(t)}-\frac{\tau}{2\varepsilon}\Phi^{T}T(U^{(t)})+\tau\omega(\hat{\mathcal{U}}-{\mathcal{U}^{(t)}})\right],
\end{equation}
where the calligraphic fonts have the meaning $\mathcal{U}=\Phi^{T}U$, ignoring the superscript for the time-step, and $\hat{\mathcal{U}}=\Phi^{T}\hat{U}$. 
Since $B=(1+c\tau)I+\varepsilon\tau\Lambda$ is a diagonal matrix with positive entries, its inverse is easy to apply. After the update, we have to project the solution back to the Gibbs simplex $\Sigma^{K}$ for the variable $U$ to reflect a probability distribution indicating to which class the $i$-th vertex belongs. In order to do this, we make use of the projection procedure in \cite{CheY11}. 

For the initialization of the segmentation problem, we first assign random values from the standard uniform distribution on $(0,1)$ to the nodes. 
Then, we project the result to the Gibbs simplex $\Sigma^{K}$ and set the values in the labeled nodes.
%
%
\end{document}